\documentclass[a4paper]{article}
\usepackage{amsfonts}
\usepackage{amsmath}
\usepackage{amssymb}
\usepackage{amsthm}
\usepackage{mathtools}
\usepackage{calc}
\usepackage{tikz-cd}
\usepackage{braket}
\usepackage{stmaryrd}
\usepackage{adjustbox}
\usepackage{multirow}

\newtheorem{assumption}{Assumption}
\newtheorem*{definition*}{Definition}

\makeatletter
\def\@crosshairs{\vbox to0pt{}}
\makeatother
\input{tcilatex}

\newcommand{\untaggedfootnote}[1]{\let\thefootnote\relax\footnote{#1}}

\def\smapinterpret{{\scalebox{0.5}[1.0]{$-$}}\hspace{-0.06cm}|}
\def\smaprepresent{|\hspace{-0.06cm}{\scalebox{0.5}[1.0]{$-$}}}
\def\mapinterpret{{\scalebox{0.5}[1.0]{$-$}}\hspace{-0.065cm}|}
\def\maprepresent{|\hspace{-0.065cm}{\scalebox{0.5}[1.0]{$-$}}}
\def\brainterpret#1{\mathinner{\langle{#1}\mapinterpret}}
\def\brarepresent#1{\mathinner{\langle{#1}\maprepresent}}

\begin{document}

\title{The Category of Linear Optical Quantum Computing}
\author{Paul McCloud \\
Department of Mathematics, University College London}
\maketitle

\begin{abstract}
This note reviews the model of computation generated by photonic circuits,
comprising edges that are traversed by photons in a single time-bin and
vertices given by idealised lossless beam splitters and phase shifters. The
circuit model is abstracted as a representation of the symmetric monoidal
category of unitary matrices on the bosonic Fock space of multimode photons.
A diagrammatic language, designed to aid with the understanding and
development of photonic algorithms, is presented that encapsulates the
category properties of this representation. As demonstrations of the
formalism, the boson sampling scheme and the protocol of Knill, Laflamme and
Milburn are developed on the single-rail photonic computer, and a
parity-based model for the qudit is investigated.

\untaggedfootnote{Author email: p.mccloud@ucl.ac.uk}
\end{abstract}

\vspace{2.75cm}

Photonic circuits perform complex mathematical operations by exploiting the
quantum mechanical properties of photons. The circuit is represented by a
graph, with directed edges traversed by photons in a fixed unit of time (the
`time-bin'), and vertices given by idealised lossless beam splitters
combined with phase shifters. Photons travel from one edge of the graph to
the next over each time-bin, proceeding from the input edges through the
internal edges to the output edges.

Each edge is associated with the state space of the photons that occupy it
within a time-bin, and each vertex is associated with an operation that maps
from the state space of its input edges to the state space of its output
edges. Circuits constructed in this way are then combined according to the
rules of symmetric monoidal categories. The circuit is thus characterised as
an operation on a quantum system with input and output state on its external
edges and internal state on its internal edges.

\begin{figure}[!t]
\centering\includegraphics[width=\linewidth]{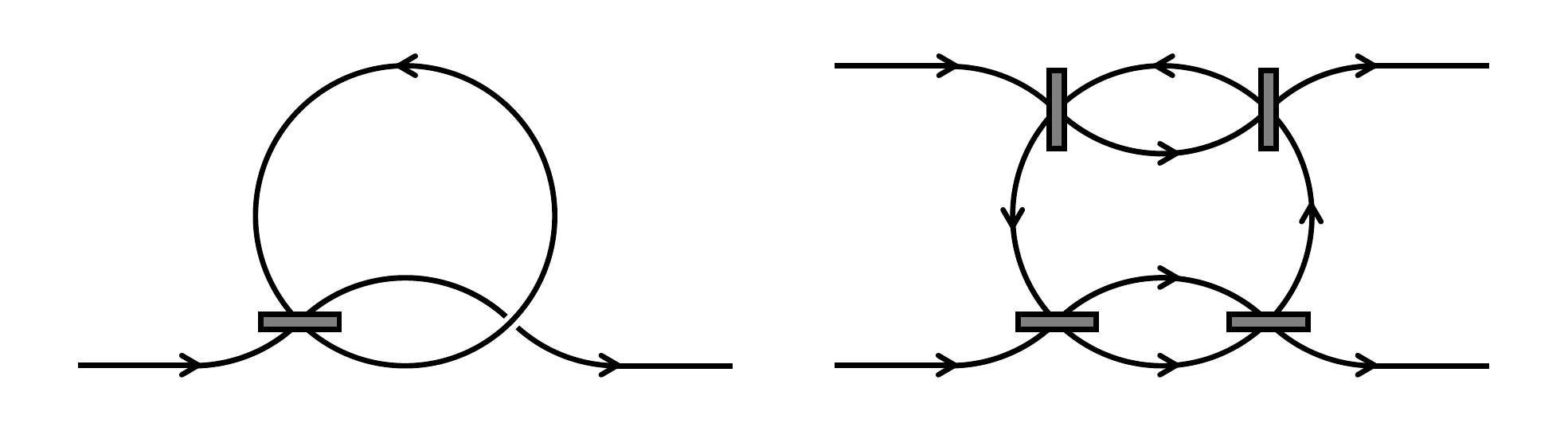}
\caption{Examples of photonic circuits. The first example has single input
and output edges, and a single internal edge that interferes with the input
via a single beam splitter. The second example has two input and output
edges, and six internal edges that interfere with the inputs and each other
via four beam splitters. Complexity is increased by adding beam splitters,
expanding the set of parameters available and enabling the construction of
sophisticated graph topologies.}
\end{figure}

Beam splitters are configured to varying degrees of interference for the
incident photons, generating transformations on the state space of the
photons that become the functions of the corresponding model of computation.
These functions are parametrised by the transmittance and reflectance of the
beam splitters and the topology of the graph. Computation on the internal
edges is seeded with photons prepared on the input edges and controlled
according to measurements on the output edges. The objective of this essay
is to identify the family of computations that can be performed on these
circuits.

\section{Circuit building}

The circuit model is abstracted in a symmetric monoidal category that
comprises systems and the operations that act between them. The rules of the
category encapsulate the commonalities of these operations across a wide
range of theoretical and physical models. Requiring no further assumptions
at this stage, the category establishes a platform for the constructions in
circuit building.

The systems considered in this essay are characterised as having external
and internal subsystems, a decomposition that constrains the methods by
which they are composed. During a single time-bin, the circuit implements an
operation:%
\begin{gather}
\includegraphics[width=\linewidth]{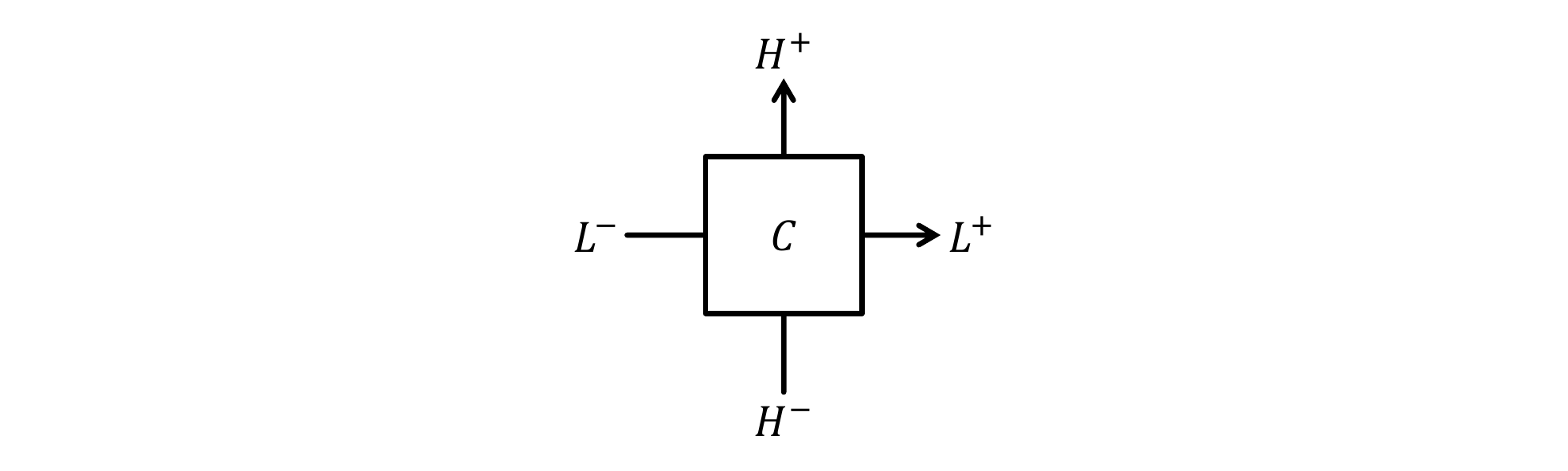}  \notag \\
C:L^{-}\otimes H^{-}\rightarrow H^{+}\otimes L^{+}
\end{gather}%
where $H^{-}$ is the input system and $H^{+}$ is the output system of the
circuit. Completing the definition, the internal system of the circuit is $%
L^{-}$ before the operation and $L^{+}$ after the operation. Systems are
represented in the diagram as legs attached to the sides of the box that
represents the operation; by convention, the leg is omitted when it
represents the empty system, the monoidal unit of the category. For
notational convenience, unitor and associator operations are omitted in the
following where they are unambiguous from the context.

As elementary examples, the identity and twist operations of the category
generate the operations:%
\begin{gather}
\includegraphics[width=\linewidth]{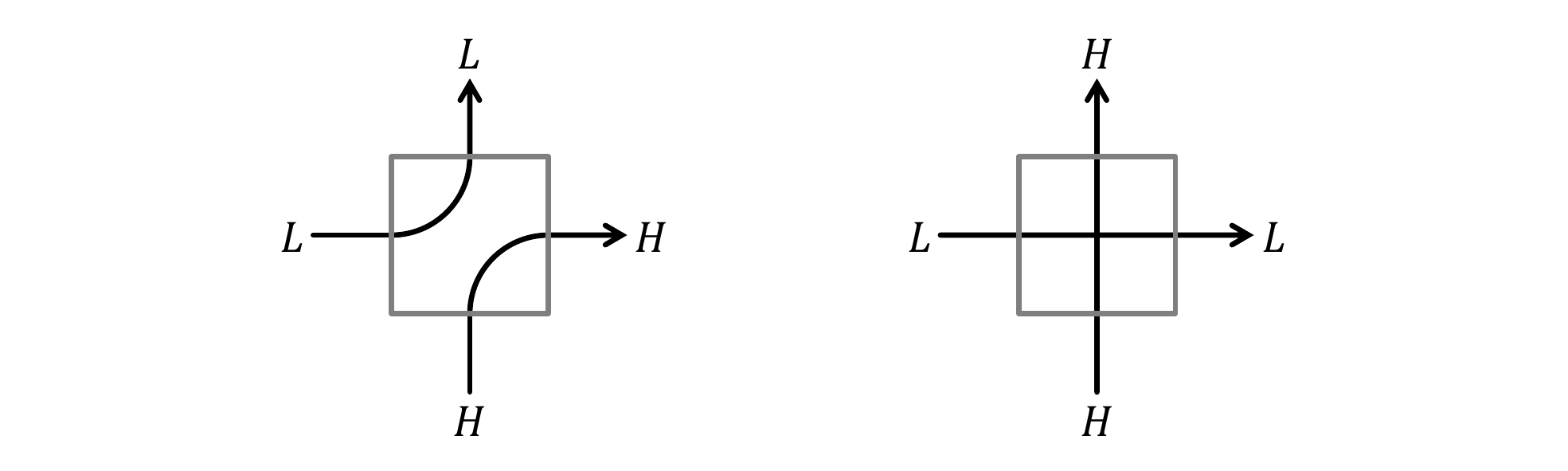}  \notag \\
\hspace{-0.2cm}\iota \otimes \iota :L\otimes H\rightarrow L\otimes H\hspace{%
1.6cm}\tau :L\otimes H\rightarrow H\otimes L
\end{gather}%
The first operation, the mirror, is given by the identity and the second
operation, the window, is given by the twist. These operations exist for any
pair of systems $H$ and $L$ and are combined to generate arbitrary system
permutations.

\begin{figure}[!p]
\begin{minipage}{\textwidth}
\begin{gather*}
\includegraphics[width=\linewidth]{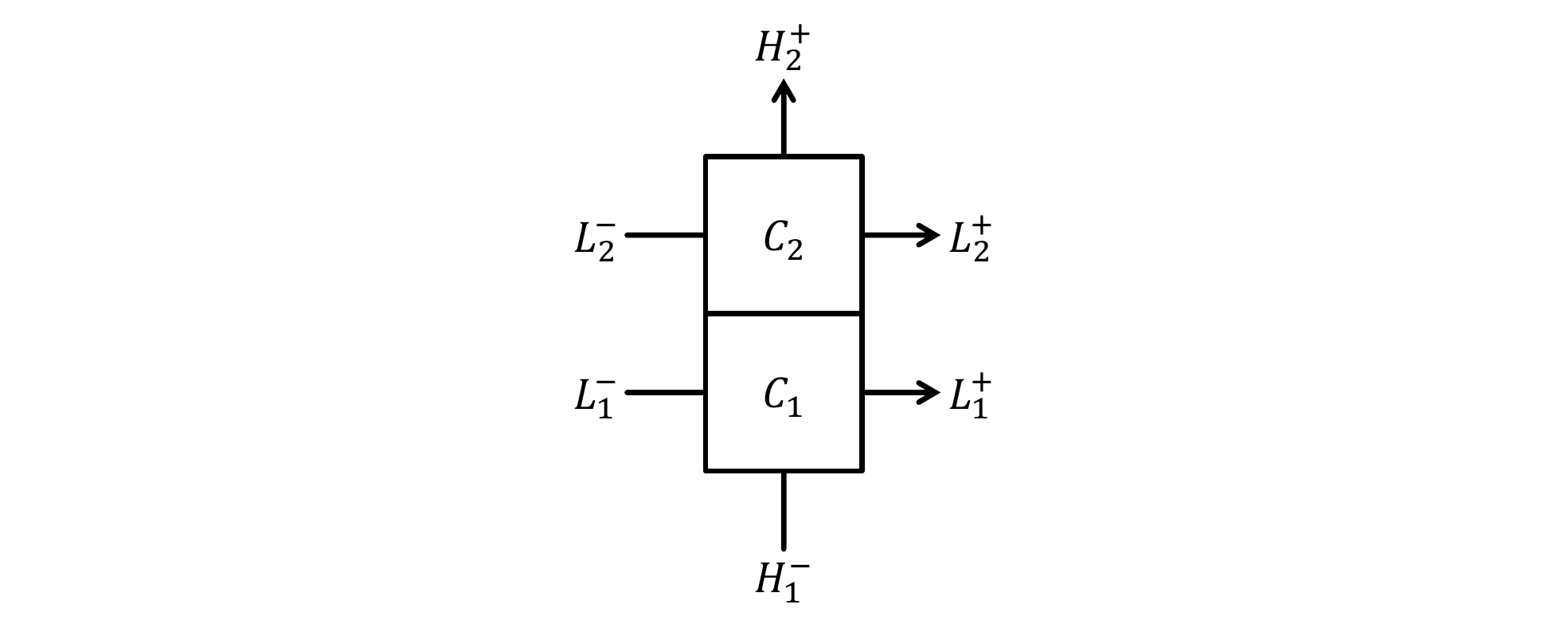} \\
C_{1}\varobar C_{2}=(\iota \otimes C_{1})\circ (C_{2}\otimes \iota ) \\
\includegraphics[width=\linewidth]{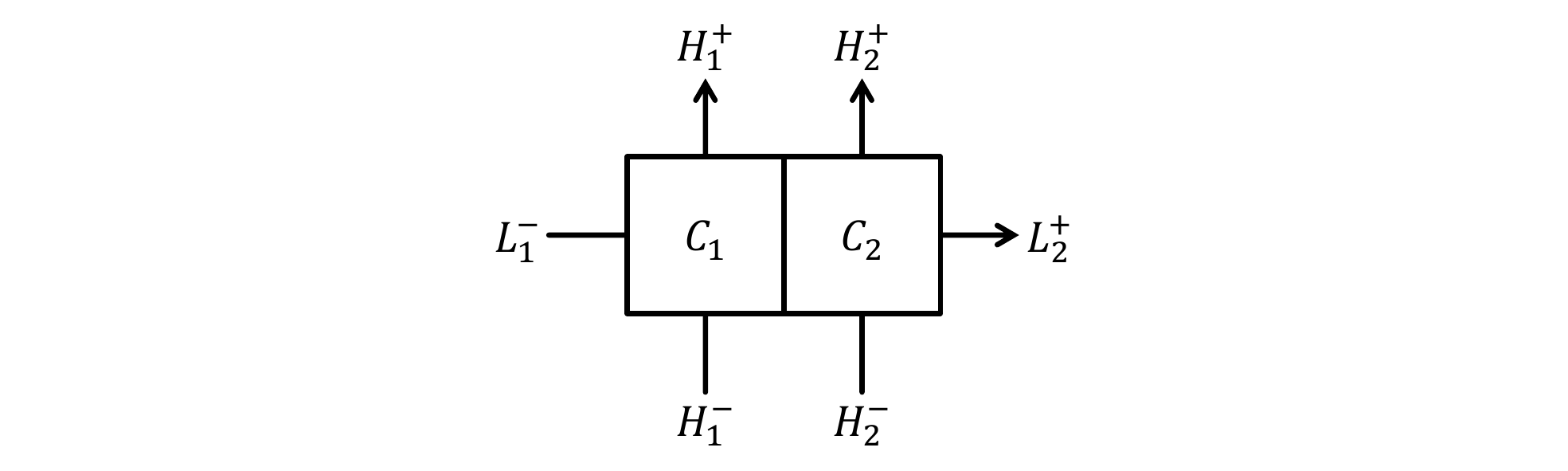} \\
C_{1}\varominus C_{2}=(C_{1}\otimes \iota )\circ (\iota \otimes C_{2})
\end{gather*}%
\caption{Circuits are constructed from elementary operations via composition. Spatial composition connects externally-compatible
circuits at their common external system. Temporal composition executes
internally-compatible circuits over consecutive time-bins.}
\end{minipage}
\newline
\newline
\newline
\begin{minipage}{\textwidth}
\begin{gather*}
\includegraphics[width=\linewidth]{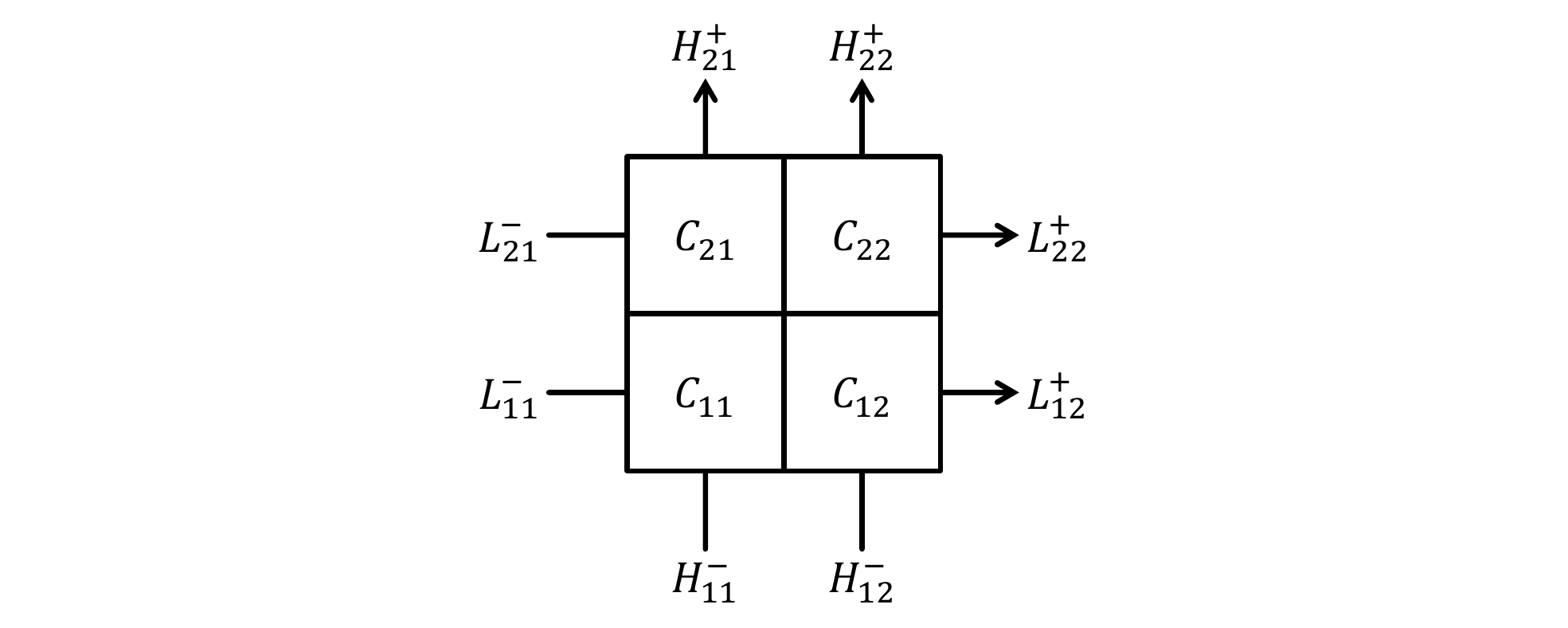} \\
(C_{11}\varobar C_{21})\varominus(C_{12}\varobar C_{22})=(C_{11}\varominus C_{12})\varobar(C_{21}\varominus C_{22})
\end{gather*}%
\caption{Spatial and temporal composition satisfy the interchange law, allowing their order of combination to be switched for compatible operations.}
\end{minipage}
\end{figure}

\begin{figure}[!p]
\begin{minipage}{\textwidth}
\centering\includegraphics[width=\linewidth]{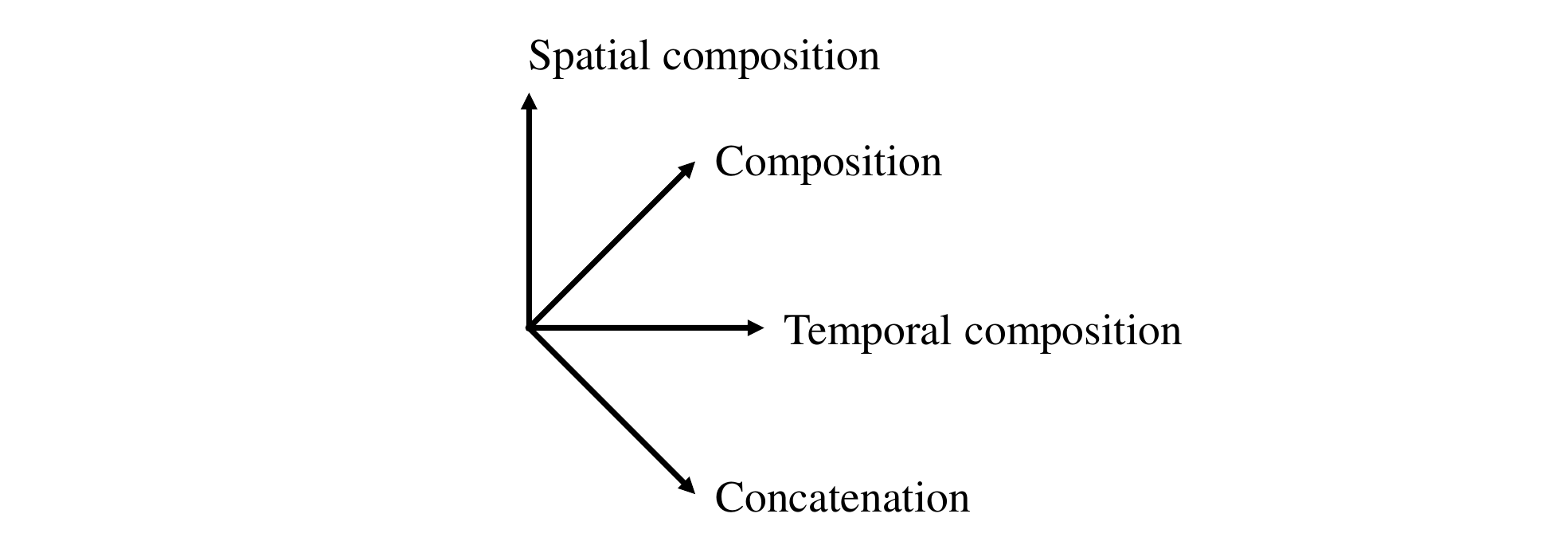}
\caption{The binary operands of spatial composition $\varobar$ and temporal composition $\varominus$ are constructed from the elementary binary operands of composition $\circ$ and concatenation $\otimes$ in the monoidal category. These operands combine along rotated axes in the diagrammatic language, offering two equivalent perspectives for the combinations of operations.}
\end{minipage}
\newline
\newline
\newline
\begin{minipage}{\textwidth}
\begin{gather*}
\includegraphics[width=\linewidth]{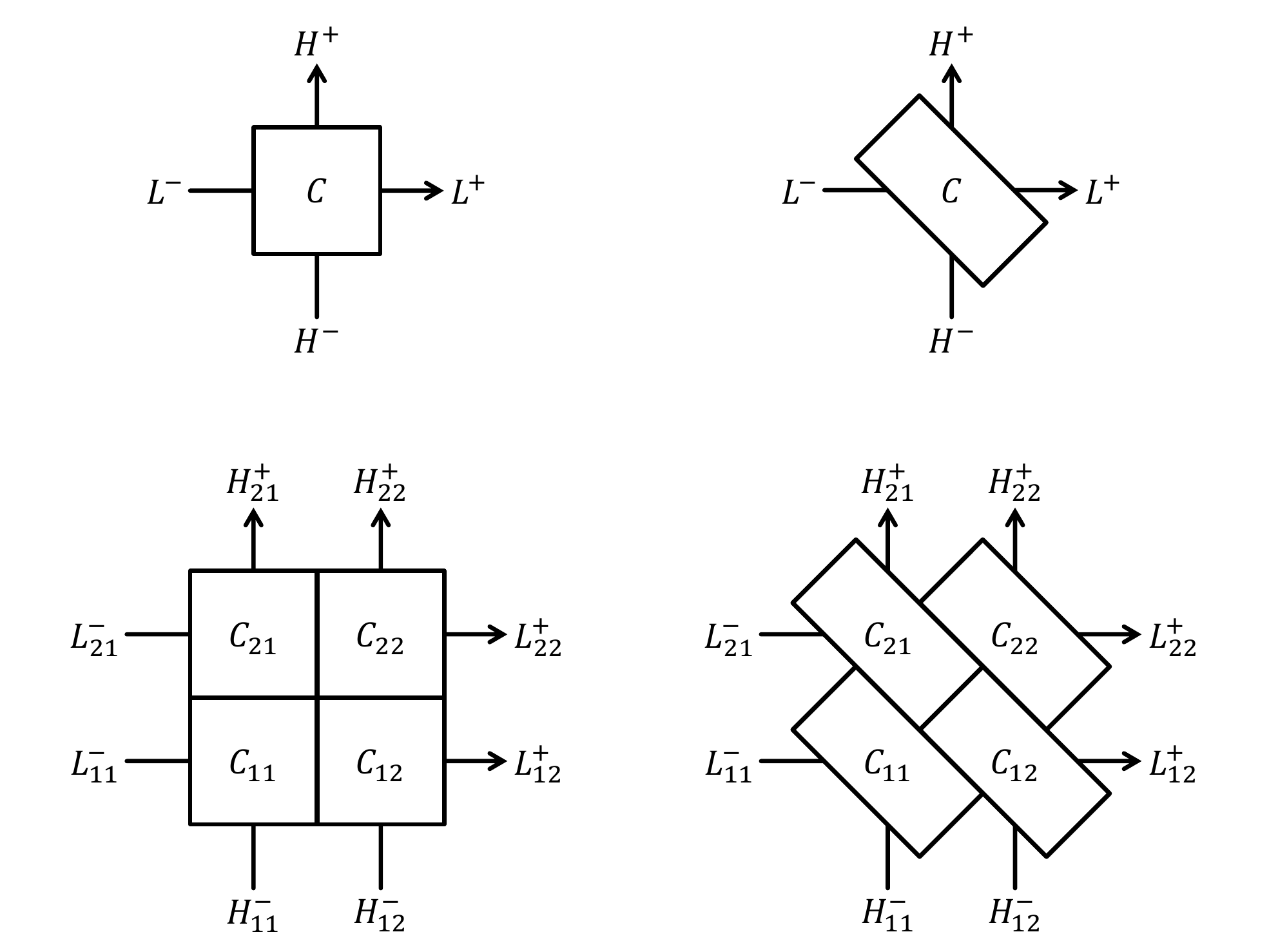} \\
(C_{11}\varobar C_{21})\varominus(C_{12}\varobar C_{22})=(C_{11}\varominus %
C_{12})\varobar(C_{21}\varominus C_{22}) \\
=(\iota \otimes C_{11}\otimes \iota )\circ (C_{21}\otimes C_{12})\circ
(\iota \otimes C_{22}\otimes \iota )
\end{gather*}%
\caption{Two perspectives on the combinations of compatible operations. On the left, operations are combined using spatial and temporal composition. On the right, operations are combined using composition and concatenation. The equivalence between these perspectives derives the interchange law.}
\end{minipage}
\end{figure}

\begin{figure}[!t]
\centering\includegraphics[width=\linewidth]{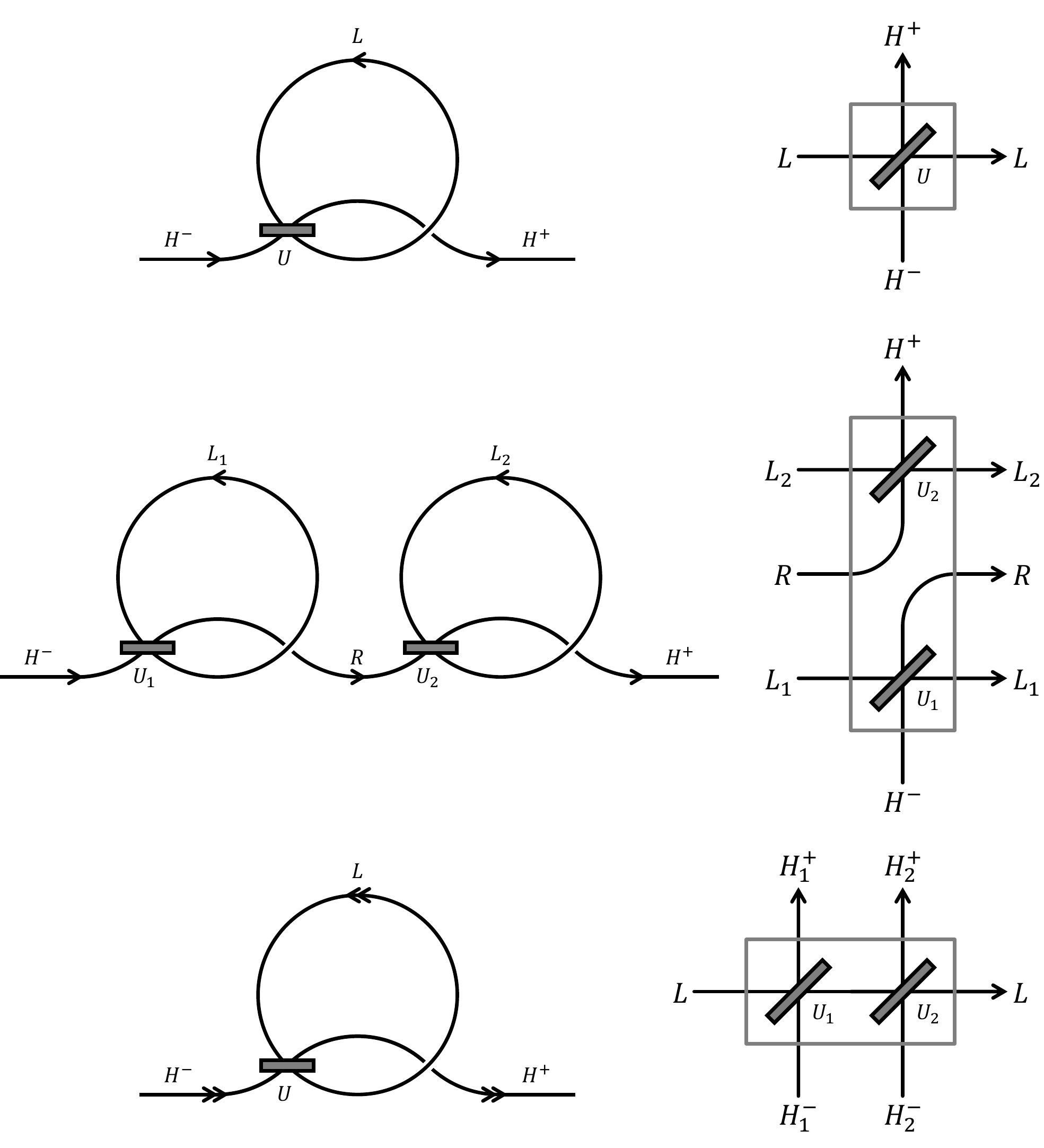}
\caption{Photonic circuit building. In these examples, the single-loop
circuit defined at the top is composed to generate two-loop circuits.
Spatial composition connects two loops at their common external edge.
Temporal composition creates a two-loop circuit by repeating the loop over
two consecutive time-bins.}
\end{figure}

\begin{figure}[!t]
\centering\includegraphics[width=\linewidth]{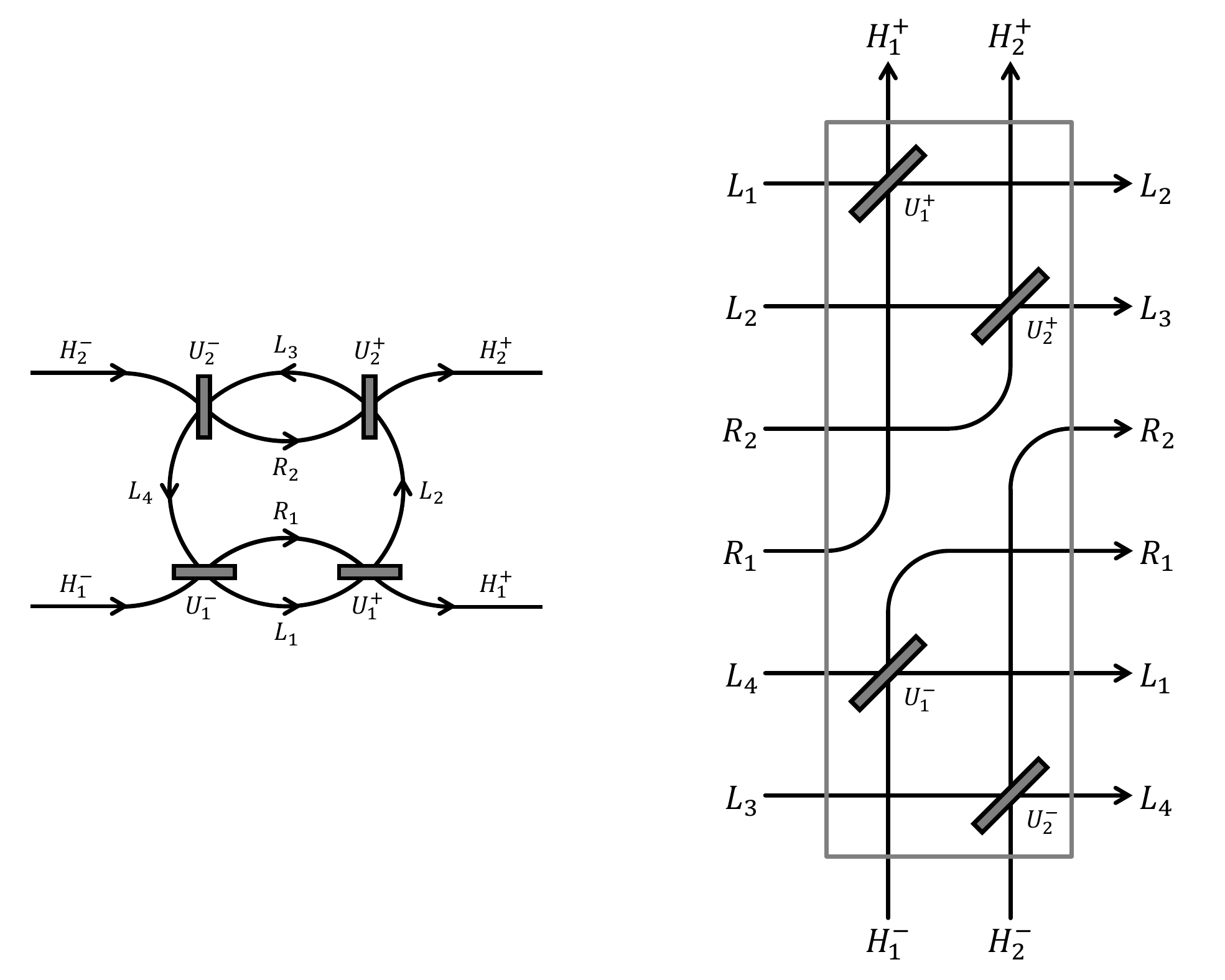}
\caption{In this example of a photonic circuit, the operation on the
external and internal edges over a single time-bin is represented by the
diagram on the right, formally constructed from mirrors, windows and beam
splitters.}
\end{figure}

Decomposing the system into its external and internal components, circuit
building is limited to two types of composition -- \emph{spatial} and \emph{%
temporal} -- each corresponding to a specific physical arrangement. These
compositions make a distinction between the external system, which can only
be spatially connected to other circuits, and the internal system, which can
only be temporally connected to other circuits.

Spatial composition is permitted for operations whose external systems are
compatible, $H_{1}^{+}=H_{2}^{-}$:%
\begin{align}
C_{1}\varobar C_{2}& =(\iota \otimes C_{1})\circ (C_{2}\otimes \iota ) \\
& :(L_{2}^{-}\otimes L_{1}^{-})\otimes H_{1}^{-}\rightarrow H_{2}^{+}\otimes
(L_{2}^{+}\otimes L_{1}^{+})  \notag
\end{align}%
This corresponds to the physical connection of two circuits at their shared
external system. Temporal composition is permitted for operations whose
internal systems are compatible, $L_{1}^{+}=L_{2}^{-}$:%
\begin{align}
C_{1}\varominus C_{2}& =(C_{1}\otimes \iota )\circ (\iota \otimes C_{2}) \\
& :L_{1}^{-}\otimes (H_{1}^{-}\otimes H_{2}^{-})\rightarrow
(H_{1}^{+}\otimes H_{2}^{+})\otimes L_{2}^{+}  \notag
\end{align}%
This corresponds to the execution of the circuit over two consecutive
time-bins. The categorical properties of these compositions are summarised
in the table:%
\begin{alignat}{4}
& & & \text{\textbf{Spatial}} & \quad & & & \text{\textbf{Temporal}}  \notag
\\
& \text{\emph{Compatibility condition:}}\quad & & H_{1}^{+}=H_{2}^{-} & & & 
& L_{1}^{+}=L_{2}^{-}  \notag \\
& \text{\emph{Input system:}}\quad & & H_{1}^{-} & & & & H_{1}^{-}\otimes
H_{2}^{-}  \notag \\
& \text{\emph{Output system:}}\quad & & H_{2}^{+} & & & & H_{1}^{+}\otimes
H_{2}^{+}  \notag \\
& \text{\emph{Initial internal system:}}\quad & & L_{2}^{-}\otimes L_{1}^{-}
& & & & L_{1}^{-}  \notag \\
& \text{\emph{Final internal system:}}\quad & & L_{2}^{+}\otimes L_{1}^{+} & 
& & & L_{2}^{+}  \notag
\end{alignat}%
Composition simplifies when the internal system of the circuit is trivial,
with spatial composition reducing to composition $C_{1}\varobar %
C_{2}=C_{1}\circ C_{2}$ and temporal composition reducing to concatenation $%
C_{1}\varominus C_{2}=C_{1}\otimes C_{2}$.

Spatial and temporal composition satisfy the interchange law:%
\begin{gather}
(C_{11}\varobar C_{21})\varominus(C_{12}\varobar C_{22})=(C_{11}\varominus %
C_{12})\varobar(C_{21}\varominus C_{22}) \\
=(\iota \otimes C_{11}\otimes \iota )\circ (C_{21}\otimes C_{12})\circ
(\iota \otimes C_{22}\otimes \iota )  \notag
\end{gather}%
While these compatible operations are composed in different ways -- spatial
then temporal on the left, and temporal then spatial on the right -- they
arrive at the same combination. This freedom is used below to decompose the
computer into operations representing preparation, computation and
measurement.

Armed with these methods of composition, the circuit model is completed by
introducing a toolkit of elementary operations. Photonic circuits introduce
the \emph{beam splitter}, which interpolates between the mirror and the
window:%
\begin{equation}
\includegraphics[width=\linewidth]{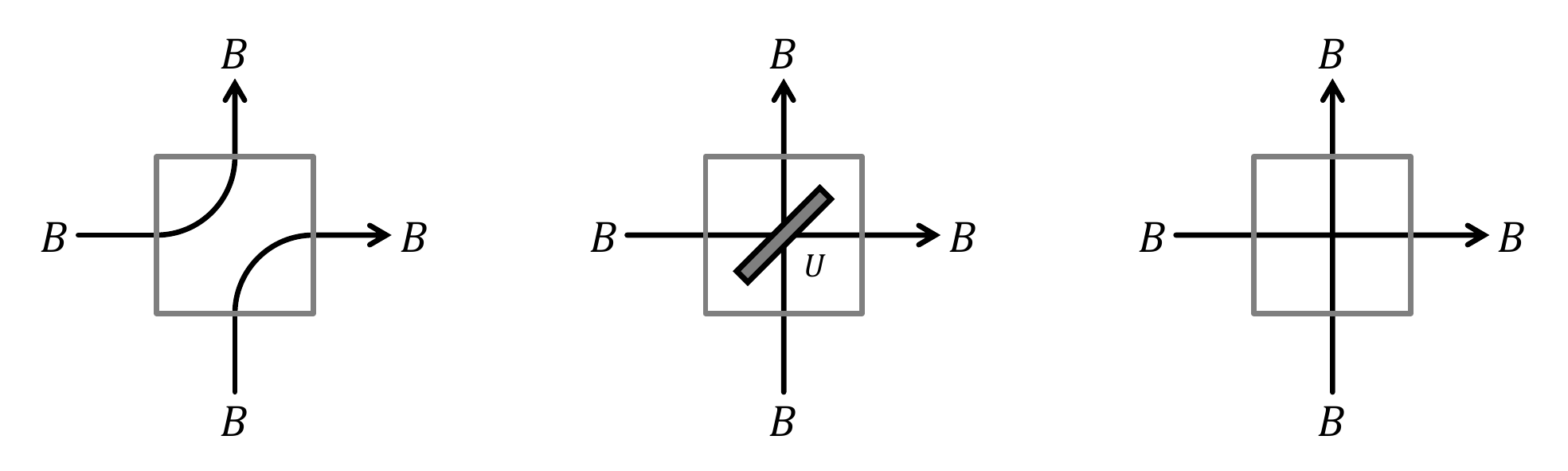}  \notag
\end{equation}%
for the photonic system $B$. This operation is novel to the photonic circuit
model, extending the elementary operations of the symmetric monoidal
category by enabling interference between systems. Quantum advantage in the
model of computation emerges from the extension of the discrete permutation
group generated by mirrors and windows to the continuous unitary group
generated by beam splitters. Classical logic gates, presented as
permutations on classical bits, are reinterpreted as unitary transformations
on quantum bits in superpositions. This argument is developed in the
following sections.

\section{Computation}

The partition of the system into its external and internal subsystems is
reflected in the model of computing. The programmer interacts with the
external system, preparing the input state and measuring the output state.
Sandwiched between these operations, computation is performed on the
internal system.

In its canonical form, execution of the circuit is comprised of three
operations representing preparation, computation and measurement:%
\begin{gather}
\includegraphics[width=\linewidth]{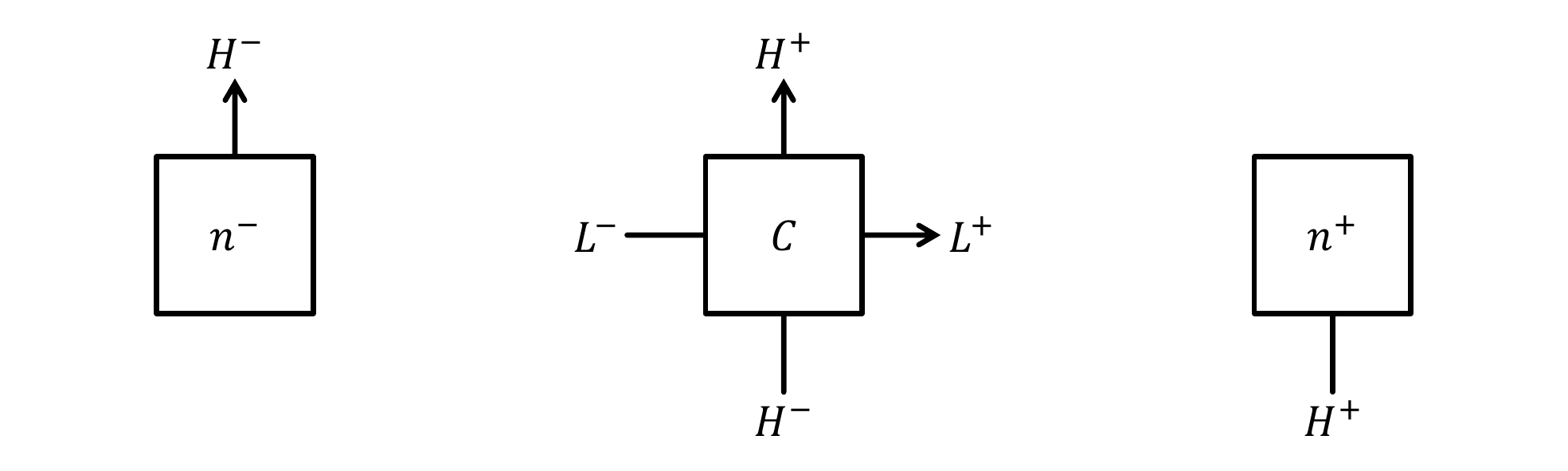}  \notag \\
\bra{n^-}:{}\rightarrow H^{-}\hspace{1.1cm}C:L^{-}\otimes H^{-}\rightarrow
H^{+}\otimes L^{+}\hspace{1.1cm}\ket{n^+}:H^{+}\rightarrow \hspace{-0.2cm}
\end{gather}%
These operations are spatially composed to generate the operation of the
circuit on the internal state over a single time-bin:%
\begin{gather}
\includegraphics[width=\linewidth]{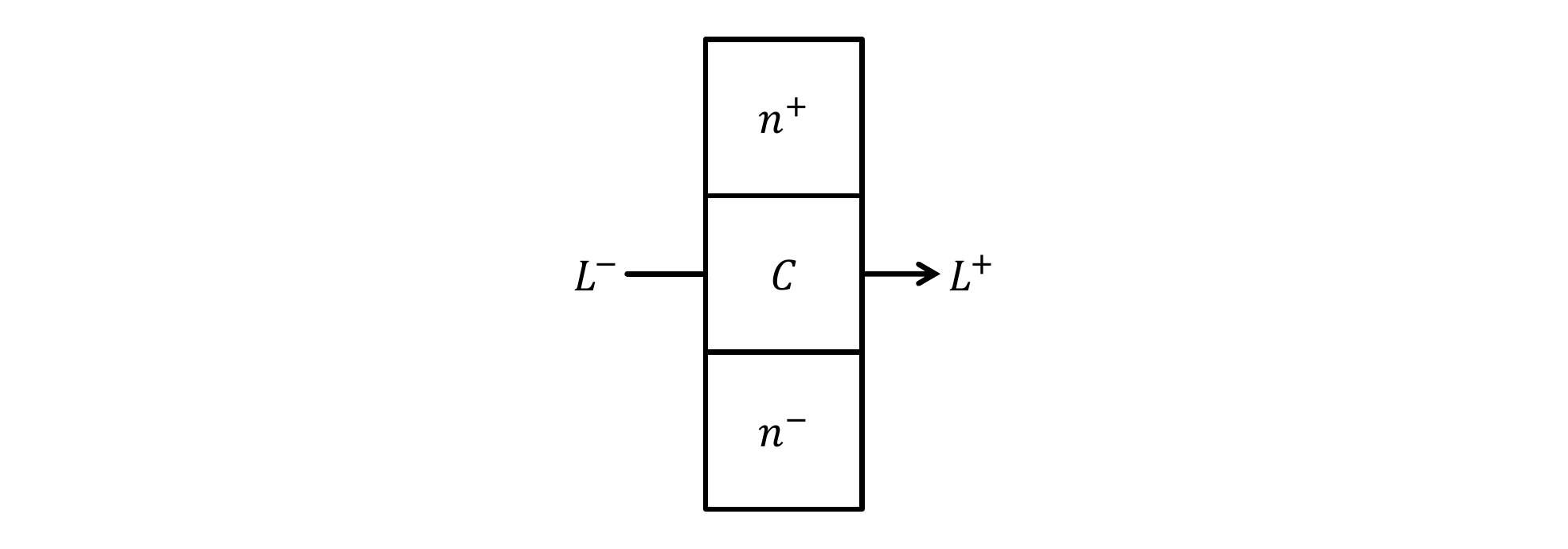}  \notag \\
\bra{n^-}\varobar C\varobar\ket{n^+}=(\iota \otimes \bra{n^-})\circ C\circ (%
\ket{n^+}\otimes \iota )
\end{gather}%
Repeated execution over multiple time-bins then generates the temporally
composed operation on the internal state:%
\begin{gather}
\includegraphics[width=\linewidth]{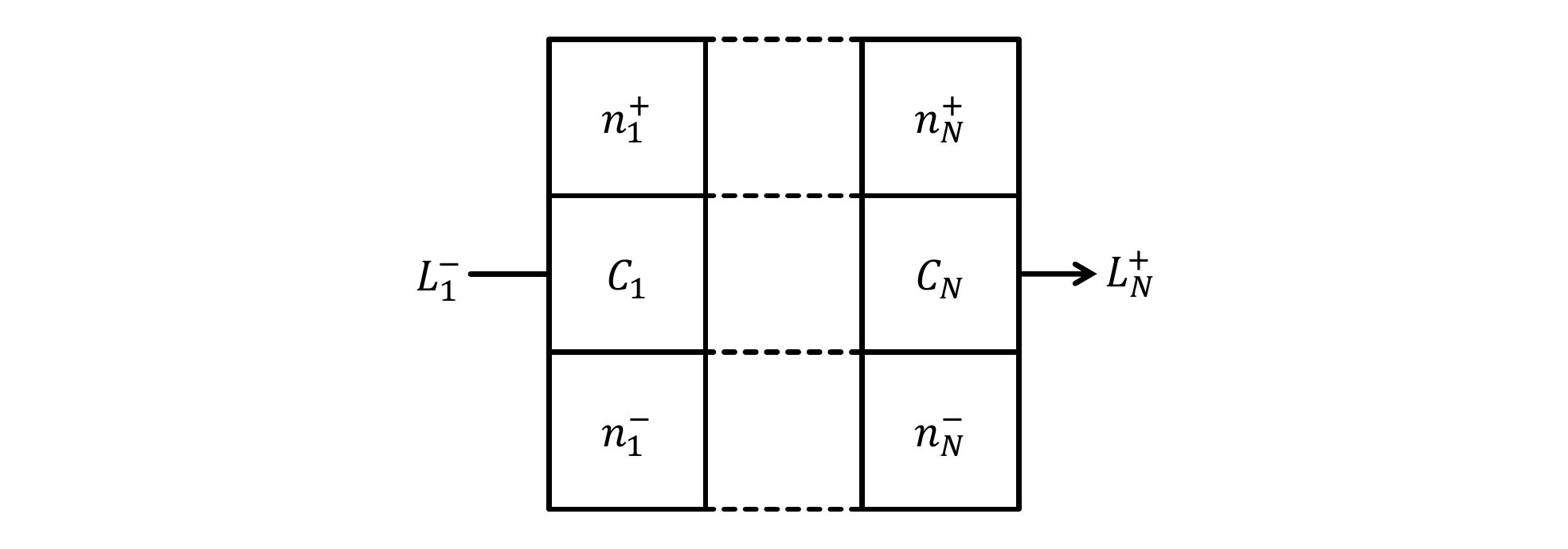}  \notag \\
(\bra{n^-_1}\varobar C_{1}\varobar\ket{n^+_1})\circ \cdots \circ (\bra{n^-_N}%
\varobar C_{N}\varobar\ket{n^+_N}) \\
=(\bra{n^-_1}\otimes \cdots \otimes \bra{n^-_N})\varobar(C_{1}\varominus%
\cdots \varominus C_{N})\varobar(\ket{n^+_1}\otimes \cdots \otimes %
\ket{n^+_N})  \notag
\end{gather}%
where the interchange law is used to express the total operation of the
circuit over multiple time-bins in canonical form, identifying the total
computation as the temporal composition of the computations in each time-bin.

Abstracted in this way, the model has the three elements essential for a
computer: an internal register where computation is performed; a means to
write data into the register; and a means to read data from the register.
For the quantum computer, the physical components that implement these
operations exploit the properties of quantum mechanics, summarised in two
postulates that describe the evolution of the quantum system and its
interaction with the external system.

To understand how these operations manifest in the quantum computer requires
specialisation to the category of quantum mechanics, wherein systems are
represented as separable Hilbert spaces and operations are bounded linear
maps between the spaces they connect, by convention acting from left to
right in the following. The monoidal unit is the trivial Hilbert space $%
\mathbb{C}$ and the monoidal product is the Hilbert space tensor product $%
\otimes $. In this category, the state $\bra{n}\in H$ is associated with two
mutually-adjoint bounded linear maps:%
\begin{align}
\bra{n}& :\mathbb{C}\rightarrow H:\lambda \mapsto \lambda \bra{n} \\
\ket{n}& :H\rightarrow \mathbb{C}:\bra{m}\mapsto \braket{m|n}  \notag
\end{align}%
utilised for preparation and measurement. Computation is then a unitary map:%
\begin{equation}
C:L^{-}\otimes H^{-}\rightarrow H^{+}\otimes L^{+}
\end{equation}%
that models the physical evolution of the system over a single time-bin.

Formulated for the application to computing, the first assumption of quantum
mechanics identifies the transformation of the internal state generated by
the quantum circuit.

\begin{assumption}[Quantum Mechanics]
If the input system is prepared in the state $\bra{n^-}\in H^{-}$ and the
output system is measured in the state $\bra{n^+}\in H^{+}$, then execution
of the circuit effects the transformation of the internal state:%
\begin{equation}
\braket{n^-|C|n^+}:=\bra{n^-}\varobar C\varobar\ket{n^+}:L^{-}\rightarrow
L^{+}
\end{equation}
\end{assumption}

\noindent This transformation combines the evolution of the internal system
and its interaction with the external system, and both these effects have
algorithmic content for the computer.

The apparatuses of the external system are associated with orthonormal bases
that diagonalise their target families of commuting observables, with the
prepared state drawn from the preparation basis and the measured state drawn
from the measurement basis. Expressing the essentially indeterminate nature
of this procedure, the fundamental asymmetry between preparation and
measurement is encapsulated in the second assumption of quantum mechanics.

\begin{assumption}[Quantum Mechanics]
If the internal system is initially in the state $\bra{m}\in L^{-}$ and the
input system is prepared in the basis state $\bra{n^-}\in H^{-}$, then the
output system is measured in the basis state $\bra{n^+}\in H^{+}$ with
probability:%
\begin{equation}
\left\Vert \bra{m}\!\braket{n^-|C|n^+}\right\Vert ^{2}/\left\Vert \bra{m}%
\right\Vert ^{2}
\end{equation}
\end{assumption}

\noindent Data is extracted from the computation by estimating these
probabilities via repeated execution of the circuit.

The computer thus comprises a circuit engineered to implement the desired
operation, coupled to apparatuses that prepare the input state and measure
the output state in their associated orthonormal bases. Central to this
calculation, the unitary map that describes the physical evolution of the
total system decomposes into blocks that individually act on the internal
state:%
\begin{equation}
C=\sum_{\bra{n^-}}\sum_{\bra{n^+}}\bra{n^+}\otimes \braket{n^-|C|n^+}\otimes %
\ket{n^-}
\end{equation}%
where the sums are over the preparation and measurement bases. Using the
effect of \emph{projective measurement}, the transformation of the internal
state executed by the circuit is non-deterministic, extracting the block $%
\braket{n^-|C|n^+}$ from the computation $C$ with row $\bra{n^-}$ selected
by the programmer but with column $\bra{n^+}$ selected at random.

\begin{figure}[!t]
\begin{minipage}{0.45\linewidth}
\centering\includegraphics[width=\linewidth]{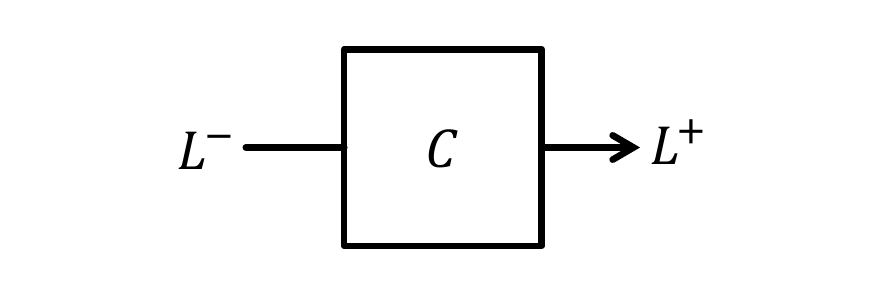}
\end{minipage}%
\begin{minipage}{0.55\linewidth}
Quantum processes are described by unitary maps, which provide the basic algorithmic components of the quantum circuit.
\end{minipage}
\noindent%
\begin{minipage}{0.45\linewidth}
\centering\includegraphics[width=\linewidth]{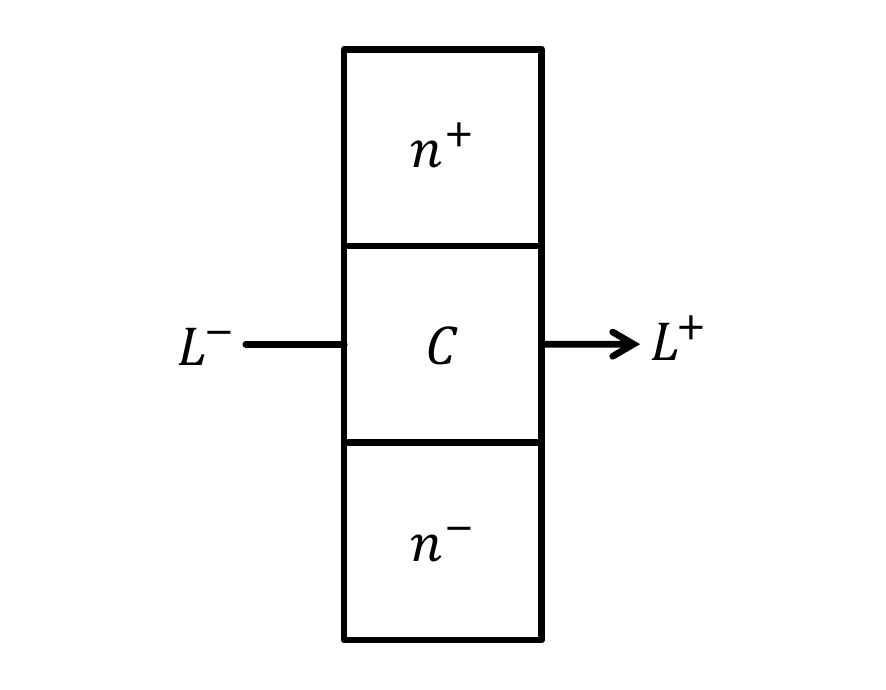}
\end{minipage}%
\begin{minipage}{0.55\linewidth}
Measurement-based computing expands the range of operations available on the circuit by exploiting projective measurement:
\begin{equation*}
\braket{n^-|C|n^+}
\end{equation*}
This is successful only when the output is measured in the required state.
\end{minipage}
\noindent%
\begin{minipage}{0.45\linewidth}
\centering\includegraphics[width=\linewidth]{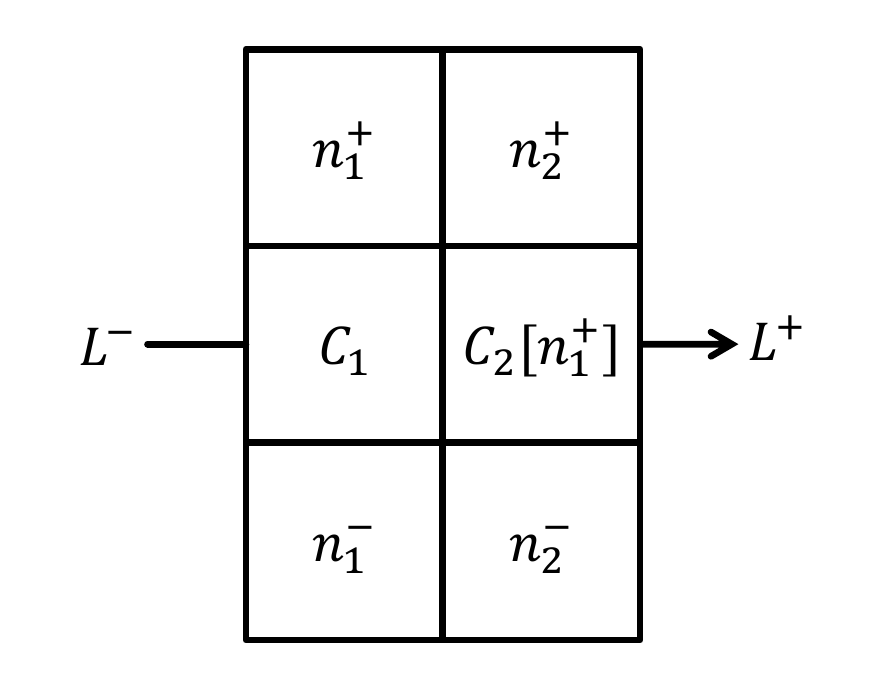}
\end{minipage}%
\begin{minipage}{0.55\linewidth}
Combined with projective measurement, the feed-forward protocol further expands the range of available operations:
\begin{equation*}
\braket{n^-_1|C_1|n^+_1}\braket{n^-_2|C_2[n^+_1]|n^+_2}
\end{equation*}
This is used to optimise the probability of successful implementation.
\end{minipage}
\caption{The physical process implements a unitary map in the quantum
computer. Projective measurement and the feed-forward protocol augment this
with the transformative effect of measurement to build algorithmic
complexity.}
\end{figure}

The algorithmic scope of the computer depends on the unitary operations that
can be implemented on the physical system and the fidelity of the
preparation and measurement apparatuses. The \emph{feed-forward protocol}
increases this scope by dynamically updating the circuit in response to
measurements in preceding time-bins. Over two time-bins, the protocol
generates an operation:%
\begin{align}
\braket{n^-_1|C_1|n^+_1}& \braket{n^-_2|C_2[n^+_1]|n^+_2} \\
& =(\bra{n^-_1}\otimes \bra{n^-_2})(C_{1}\varominus C_{2})(\ket{n^+_1}%
\otimes \ket{n^+_2})  \notag
\end{align}%
The interchange law derives the total computation as a temporal composition,
using orthonormality of the measurement basis to extend the definition:%
\begin{equation}
C_{1}\varominus C_{2}=(C_{1}\otimes \iota )\sum_{\bra{n^+_1}}(\ket{n^+_1}%
\bra{n^+_1}\otimes C_{2}[n_{1}^{+}])
\end{equation}%
While the practical challenges of dynamic updating in physical circuits
should not be underestimated, this extended protocol for temporal
composition greatly expands the algorithmic range.

The lack of determinism is a problem with an approach that utilises
projective measurement. For a computer engineered to a specific task, the
desired operation may not be implemented in all possible outcomes, with
success or failure indicated by the measurement of the output state.
Fortunately quantum mechanics presents a practical solution to this
conundrum. Non-deterministic protocols can be used to prepare clusters of
entangled states in advance of computation. Measurement-based computing then
proceeds via targeted measurement of the cluster state resource, using the
feed-forward protocol to generate the desired operation on the internal
state.

\section{Photonic operations}

With the platform for circuit building established as the symmetric monoidal
category of quantum systems, this section develops the family of operations
generated by linear optical components acting on photons. Conveniently,
these operations are described by a monoidal functor that represents unitary
matrices on multimode bosonic Fock spaces.

The state space of indistinguishable photons in a single mode is the bosonic
Fock space $B$, with vacuum state $\bra{0}$ and orthonormal computation
basis given by states of the form:%
\begin{equation}
\bra{n}=\frac{1}{\sqrt{n!}}\bra{0}\!a^{n}
\end{equation}%
where the operation $a:B\rightarrow B$ is the photon creation operator on
the state space. The $N$-mode state space $B[N]$ is the $N$th tensor power
of $B$, and the creation and annihilation operators $a_{M}$ and $a_{M}^{\ast
}$ for the $M$-mode are:%
\begin{align}
\bra{n_1\cdots n_N}\!a_{M}& =\sqrt{n_{M}+1}\,\bra{n_1\cdots(n_M+1)\cdots n_N}
\\
\bra{n_1\cdots n_N}\!a_{M}^{\ast }& =\sqrt{n_{M}}\,\bra{n_1\cdots(n_M-1)%
\cdots n_N}  \notag
\end{align}%
The number operator $\hat{n}_{M}$ for the $M$-mode is then:%
\begin{equation}
\hat{n}_{M}=a_{M}a_{M}^{\ast }-1
\end{equation}%
and the computation basis of the $N$-mode state space comprises the
eigenstates for its family of number operators.

In the linear optical model, a computation in the photonic circuit is
determined by its commutation relations with the creation operators on the
state spaces of its edges. The operation:%
\begin{equation}
B[U]:H^{-}\rightarrow H^{+}
\end{equation}%
with state spaces $H^{-}=H^{+}=B[N]$ is generated by a unitary $N$-matrix $U$
via the commutation relations:%
\begin{equation}
a_{M}B[U]=B[U]\sum_{L=1}^{N}U_{ML}a_{L}
\end{equation}%
These relations are sufficient to determine the operation on the basis
states up to an overall normalisation:%
\begin{align}
\bra{n_1\cdots n_N}\!B[U]& =(\frac{1}{\sqrt{n_{1}!\cdots n_{N}!}}%
\bra{0\cdots 0}\!\prod_{M=1}^{N}a_{M}^{n_{M}})B[U] \\
& =\frac{1}{\sqrt{n_{1}!\cdots n_{N}!}}\bra{0\cdots 0}\!\prod_{M=1}^{N}(%
\sum_{L=1}^{N}U_{ML}a_{L})^{n_{M}}  \notag
\end{align}%
where the definition is completed with the normalisation:%
\begin{equation}
\bra{0\cdots 0}\!B[U]=\bra{0\cdots 0}
\end{equation}
on the vacuum state. Expanding the brackets and applying the creation
operators to the vacuum state, this generates an operation that preserves
the total photon number:%
\begin{equation}
\bra{n^-_1\cdots n^-_N}\!B[U]=\sum_{\substack{ n_{1}^{+}+\cdots +n_{N}^{+}= 
\\ \hfill \mathstrut n_{1}^{-}+\cdots +n_{N}^{-}}}\braket{n^-_1\cdots
n^-_N|B[U]|n^+_1\cdots n^+_N}\bra{n^+_1\cdots n^+_N}
\end{equation}%
where the matrix elements of $B[U]$ on the eigenspace of total photon number 
$n$ are linear combinations of products of $n$ elements from $U$.

The map $B$ as defined here is a monoidal functor. The source category of
the functor comprises the strictly positive integers with unitary matrices
in the corresponding dimensions. The target category of the functor
comprises the multimode bosonic Fock spaces with the unitary maps they
support. Monoidal functoriality for objects is the property:%
\begin{equation}
B[N_{1}+N_{2}]=B[N_{1}]\otimes B[N_{2}]
\end{equation}%
while monoidal functoriality for morphisms is the properties:%
\begin{align}
B[U_{1}\oplus U_{2}]& =B[U_{1}]\otimes B[U_{2}] \\
B[U_{1}U_{2}]& =B[U_{1}]B[U_{2}]  \notag
\end{align}%
for compatible unitary matrices. Photonic circuits thus create a
representation of the category of finite-dimensional unitary matrices as
unitary operations on the multimode bosonic Fock spaces of photons.

\begin{figure}[!t]
\centering\includegraphics[width=\linewidth]{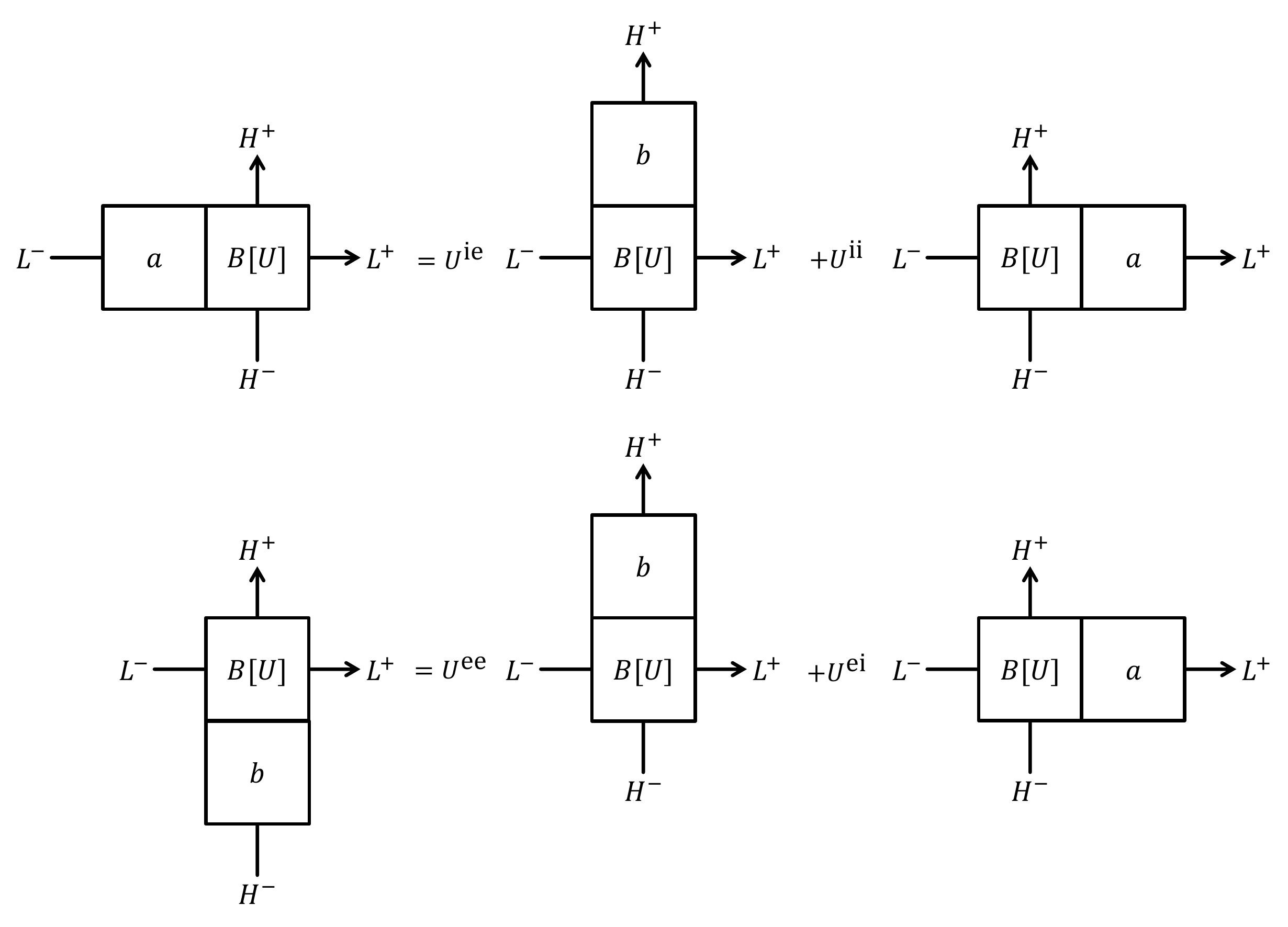}
\caption{The commutation relations between a circuit $B[U]$ and the photon
creation operators $a$ on its internal edges and $b$ on its external edges
are generated from a unitary matrix $U$.}
\end{figure}

\begin{figure}[!p]
\begin{minipage}{\textwidth}
\centering\includegraphics[width=\linewidth]{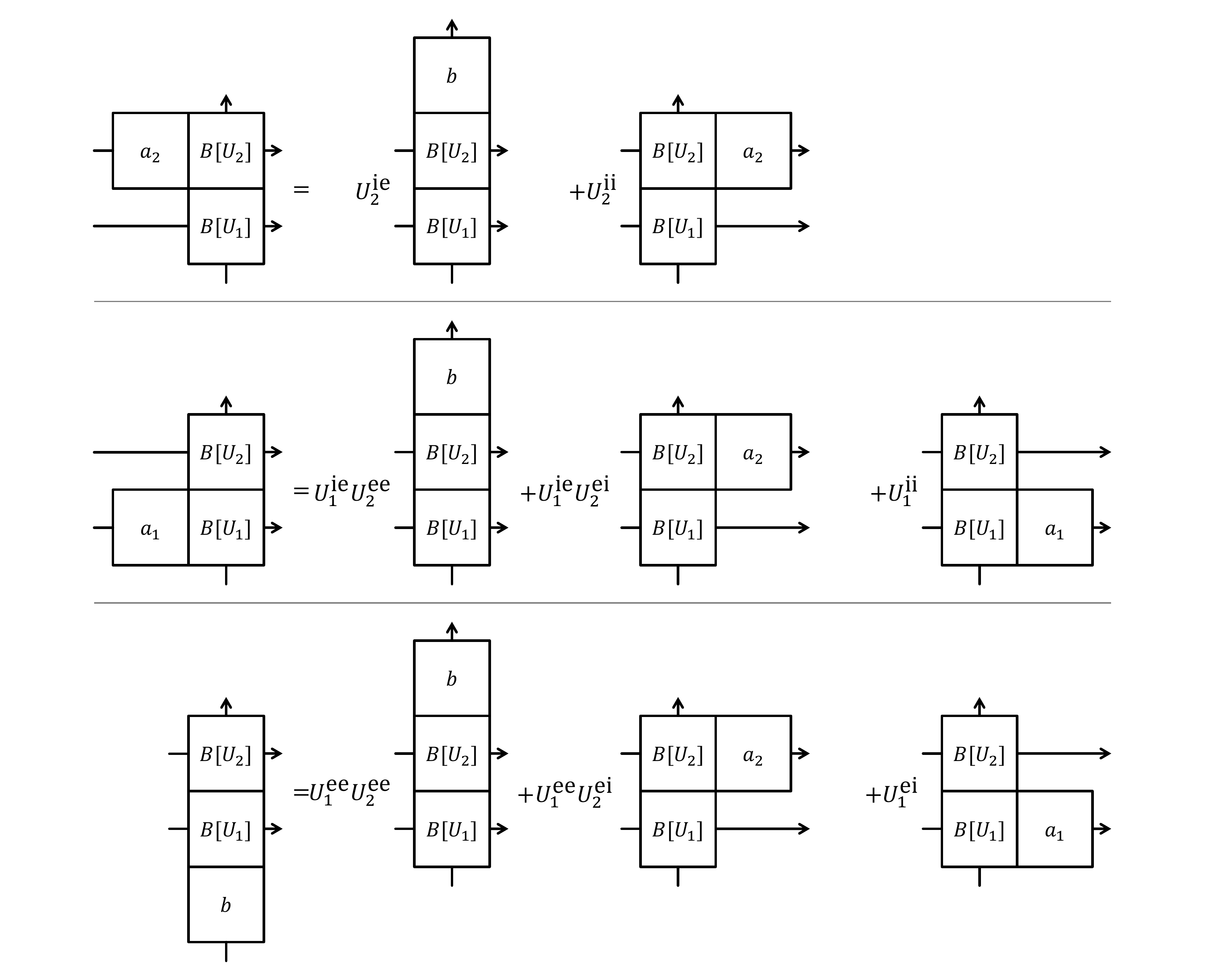}
\caption{The unitary matrix that generates spatial composition is identified
by commuting the photon creation operators through its two circuits.}
\end{minipage}
\newline
\newline
\newline
\begin{minipage}{\textwidth}
\centering\includegraphics[width=\linewidth]{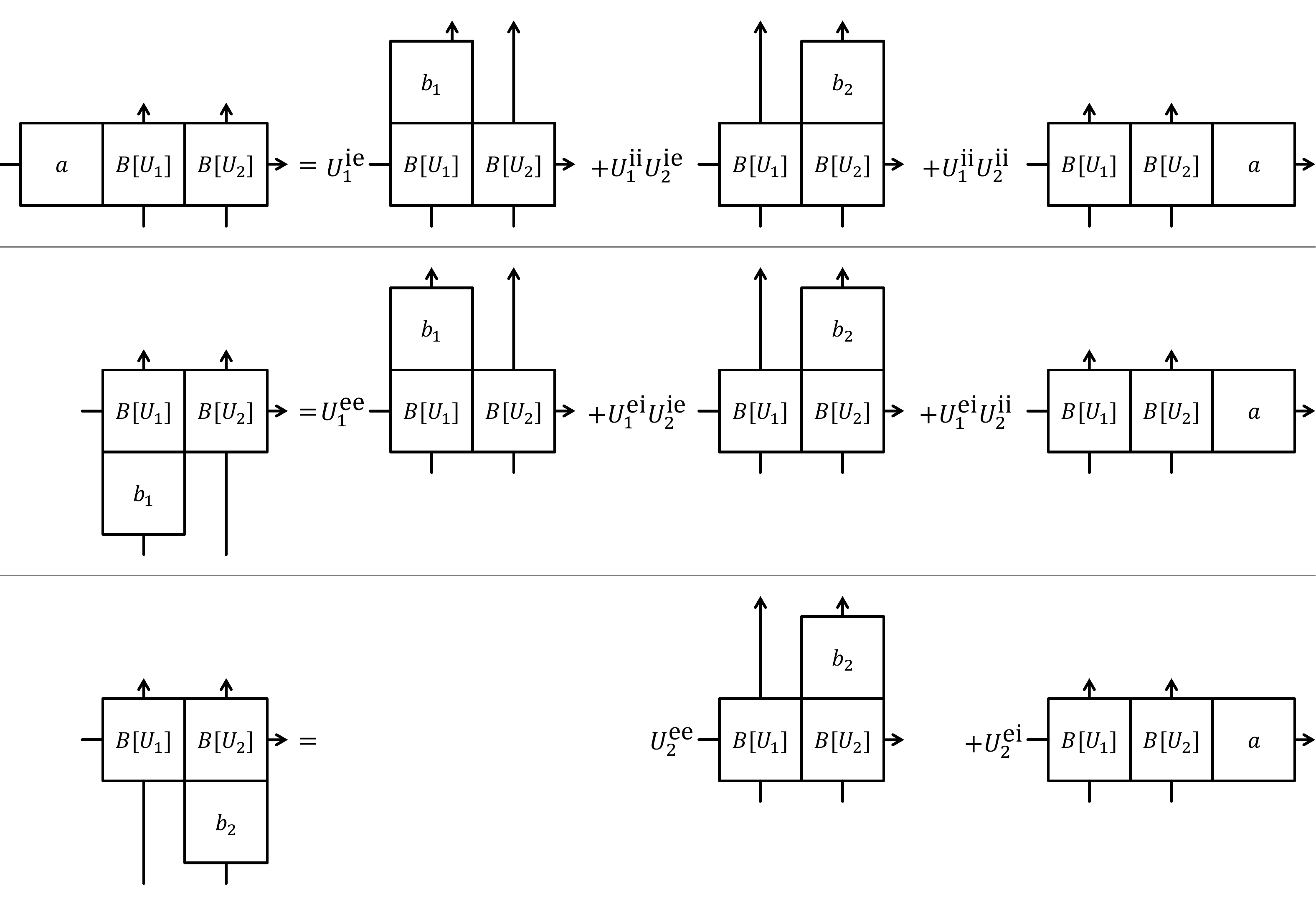}
\caption{The unitary matrix that generates temporal composition is
identified by commuting the photon creation operators through its two
circuits.}
\end{minipage}
\end{figure}

For the operation:%
\begin{equation}
B[U]:L^{-}\otimes H^{-}\rightarrow H^{+}\otimes L^{+}
\end{equation}%
whose state spaces decompose into internal state spaces $L^{-}=L^{+}=B[M]$
and external state spaces $H^{-}=H^{+}=B[N]$, the unitary matrix $U$ that
generates the operation decomposes into blocks:%
\begin{equation}
U=%
\begin{bmatrix}
U^{\mathrm{ie}} & U^{\mathrm{ii}} \\ 
U^{\mathrm{ee}} & U^{\mathrm{ei}}%
\end{bmatrix}%
\end{equation}%
where the superscript e indicates external edges and the superscript i
indicates internal edges. The functorial nature of the map from the matrix $%
U $ to the operation $B[U]$ that it generates allows the calculus of circuit
building to be expressed by constructions within the category of
finite-dimensional unitary matrices. This simplifies the analysis of
circuits.

Consider the pair of operations:%
\begin{align}
B[U_{1}]& :L_{1}^{-}\otimes H_{1}^{-}\rightarrow H_{1}^{+}\otimes L_{1}^{+}
\\
B[U_{2}]& :L_{2}^{-}\otimes H_{2}^{-}\rightarrow H_{2}^{+}\otimes L_{2}^{+} 
\notag
\end{align}%
Compositions of these operations are efficiently described by the unitary
matrices in their commutation relations. The operations are spatially
compatible when $N_{1}=N_{2}$, in which case the unitary matrix that
generates the spatial composition is:%
\begin{align}
(1\oplus U_{1})(U_{2}\oplus 1)& =%
\begin{bmatrix}
1 & 0 & 0 \\ 
0 & U_{1}^{\mathrm{ie}} & U_{1}^{\mathrm{ii}} \\ 
0 & U_{1}^{\mathrm{ee}} & U_{1}^{\mathrm{ei}}%
\end{bmatrix}%
\begin{bmatrix}
U_{2}^{\mathrm{ie}} & U_{2}^{\mathrm{ii}} & 0 \\ 
U_{2}^{\mathrm{ee}} & U_{2}^{\mathrm{ei}} & 0 \\ 
0 & 0 & 1%
\end{bmatrix}
\\
& =%
\begin{bmatrix}
U_{2}^{\mathrm{ie}} & U_{2}^{\mathrm{ii}} & 0 \\ 
U_{1}^{\mathrm{ie}}U_{2}^{\mathrm{ee}} & U_{1}^{\mathrm{ie}}U_{2}^{\mathrm{ei%
}} & U_{1}^{\mathrm{ii}} \\ 
U_{1}^{\mathrm{ee}}U_{2}^{\mathrm{ee}} & U_{1}^{\mathrm{ee}}U_{2}^{\mathrm{ei%
}} & U_{1}^{\mathrm{ei}}%
\end{bmatrix}
\notag
\end{align}%
The operations are temporally compatible when $M_{1}=M_{2}$, in which case
the unitary matrix that generates the temporal composition is:%
\begin{align}
(U_{1}\oplus 1)(1\oplus U_{2})& =%
\begin{bmatrix}
U_{1}^{\mathrm{ie}} & U_{1}^{\mathrm{ii}} & 0 \\ 
U_{1}^{\mathrm{ee}} & U_{1}^{\mathrm{ei}} & 0 \\ 
0 & 0 & 1%
\end{bmatrix}%
\begin{bmatrix}
1 & 0 & 0 \\ 
0 & U_{2}^{\mathrm{ie}} & U_{2}^{\mathrm{ii}} \\ 
0 & U_{2}^{\mathrm{ee}} & U_{2}^{\mathrm{ei}}%
\end{bmatrix}
\\
& =%
\begin{bmatrix}
U_{1}^{\mathrm{ie}} & U_{1}^{\mathrm{ii}}U_{2}^{\mathrm{ie}} & U_{1}^{%
\mathrm{ii}}U_{2}^{\mathrm{ii}} \\ 
U_{1}^{\mathrm{ee}} & U_{1}^{\mathrm{ei}}U_{2}^{\mathrm{ie}} & U_{1}^{%
\mathrm{ei}}U_{2}^{\mathrm{ii}} \\ 
0 & U_{2}^{\mathrm{ee}} & U_{2}^{\mathrm{ei}}%
\end{bmatrix}
\notag
\end{align}%
These expressions describe the mathematics of photonic circuit building in
terms of the generating unitary matrices.

The basic operation in photonic circuit building is the beam splitter:%
\begin{gather}
\includegraphics[width=\linewidth]{Diagram-MirrorSplitterWindow.pdf}  \notag
\\
I=%
\begin{bmatrix}
1 & 0 \\ 
0 & 1%
\end{bmatrix}%
\hspace{0.8cm}U=e^{i\gamma }%
\begin{bmatrix}
e^{-i\rho }\sin [\theta ] & e^{-i\tau }\cos [\theta ] \\ 
e^{i\tau }\cos [\theta ] & -e^{i\rho }\sin [\theta ]%
\end{bmatrix}%
\hspace{0.7cm}X=%
\begin{bmatrix}
0 & 1 \\ 
1 & 0%
\end{bmatrix}%
\hspace{-0.5cm}
\end{gather}%
The beam splitter has transmittance $\cos [\theta ]^{2}$ and reflectance $%
\sin [\theta ]^{2}$. Combined with phase shifters, the beam splitter
introduces a global phase $\gamma $ together with phase difference $\tau $
between the transmitted photons and phase difference $\rho $ between the
reflected photons. This creates any generating unitary matrix in two
dimensions.

Elementary examples of configurations for the beam splitter include the
mirror, with $\theta =\gamma =\rho =\pi /2$, and the window, with $\theta
=\gamma =\tau =0$. More generally, the operation is used to interfere the
input photons and generate mixed states on the output edges. Beam splitters
are then combined via spatial and temporal composition to create any
generating unitary matrix in $N$ dimensions.

Circuits constructed from this optical component include the single-rail
photonic computer, a simple architecture which nonetheless implements a wide
range of photonic operations.

\begin{figure}[!t]
\centering\includegraphics[width=\linewidth]{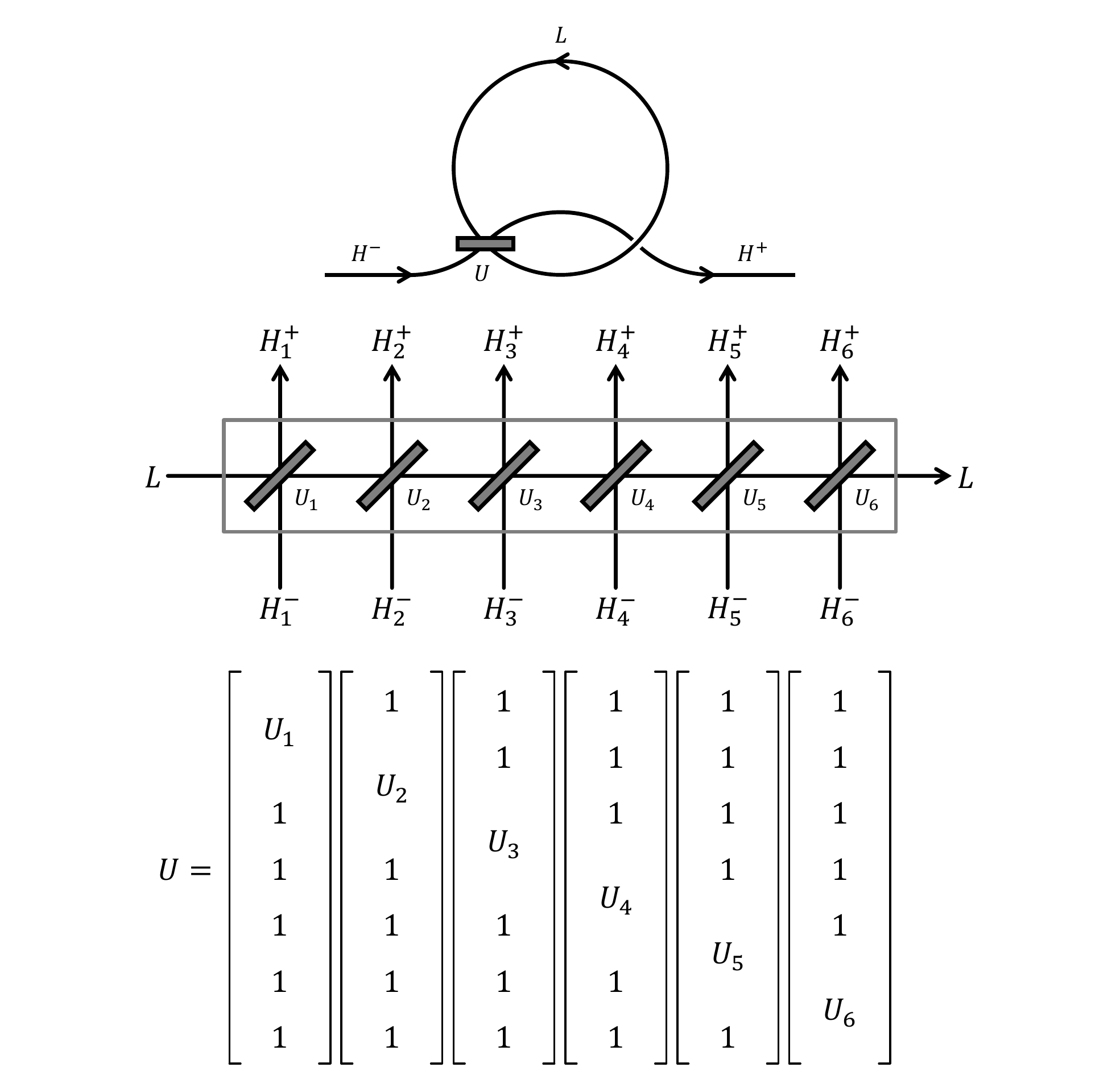}
\caption{The single-loop photonic computer. The physical circuit comprises a
single loop executed over $N_H$ consecutive time-bins ($N_H=6$ in this
example). The generating unitary matrix of the operation is the product of $%
N_H$ block-diagonal unitary matrices, represented here as the column vectors
of their diagonal blocks.}
\end{figure}

\paragraph{The single-loop photonic computer.}

The simplest non-trivial photonic circuit is the single-loop circuit, with
one input and output edge interfering with an internal edge via a
configurable beam splitter.

The beam splitter is configured to the reflectance and phase shifts of a
unitary $2$-matrix $U$. When the input is prepared in the basis state $%
\bra{n^-}\in H^{-}$ and the output is measured in the basis state $\bra{n^+}%
\in H^{+}$, conservation of photon number implies that the action on the
internal state is:%
\begin{equation}
\bra{m^-}\!\braket{n^-|B[U]|n^+}=\braket{m^-n^-|B[U]|n^+m^+}\bra{m^+}
\end{equation}%
where:%
\begin{equation}
m^{+}=m^{-}+n^{-}-n^{+}
\end{equation}%
For such a simple circuit, this operation is surprisingly complex. The
computer does not mix internal basis states but does transform the
coefficients, with amplitudes determined from the commutation relation as:%
\begin{align}
& \braket{m^-n^-|B[U]|n^+m^+}= \\
& \quad \sqrt{\frac{n^{+}!m^{+}!}{m^{-}!n^{-}!}}e^{i((m^{-}+n^{-})\gamma
-(m^{-}-n^{+})\tau +(n^{-}-n^{+})\rho )}\times  \notag \\
& \quad (-)^{\min [n^{-},m^{+}]}\cos [\theta ]^{|m^{-}-n^{+}|}\sin [\theta
]^{|n^{-}-n^{+}|}\times  \notag \\
& \quad \sum_{\eta =0}^{\kappa }\binom{m^{-}}{\min [m^{-},n^{+}]-\eta }%
\binom{n^{-}}{\min [n^{-},n^{+}]-(\kappa -\eta )}(-)^{\eta }\cos [\theta
]^{2\eta }\sin [\theta ]^{2(\kappa -\eta )}  \notag
\end{align}%
where:%
\begin{equation}
\kappa =\min [m^{-},n^{+}]+\min [n^{-},n^{+}]-n^{+}
\end{equation}%
These amplitudes demonstrate the trigonometric nature of interference
between the photons incident on the beam splitter.

With careful alignment of the angles in successive time-bins, interference
and projective measurement are exploited to create useful operations on the
single-loop circuit. Since this depends on the output photon count, the
target operation is implemented non-deterministically with success rate
given by the squared modulus of the amplitude.

As an example, implementation of the non-linear sign gate on the single-loop
photonic computer uses the following diagonal cases:%
\begin{align}
\bra{m}\!\braket{0|B[U]|0}& =e^{im(\gamma -\tau )}\cos [\theta ]^{m}\bra{m}
\\
\bra{m}\!\braket{1|B[U]|1}& =-e^{i((m+1)\gamma -(m-1)\tau )}\cos [\theta
]^{m-1}(m-(1+m)\cos [\theta ]^{2})\bra{m}  \notag
\end{align}%
to change the sign of the internal basis state conditional on its photon
count. This operation is integral to the Knill, Laflamme and Milburn
implementation of the controlled-Z gate, which enables universal quantum
computing on the single-rail circuit. Details of this implementation are
presented later in the essay.

\begin{figure}[!p]
\centering\includegraphics[width=\linewidth]{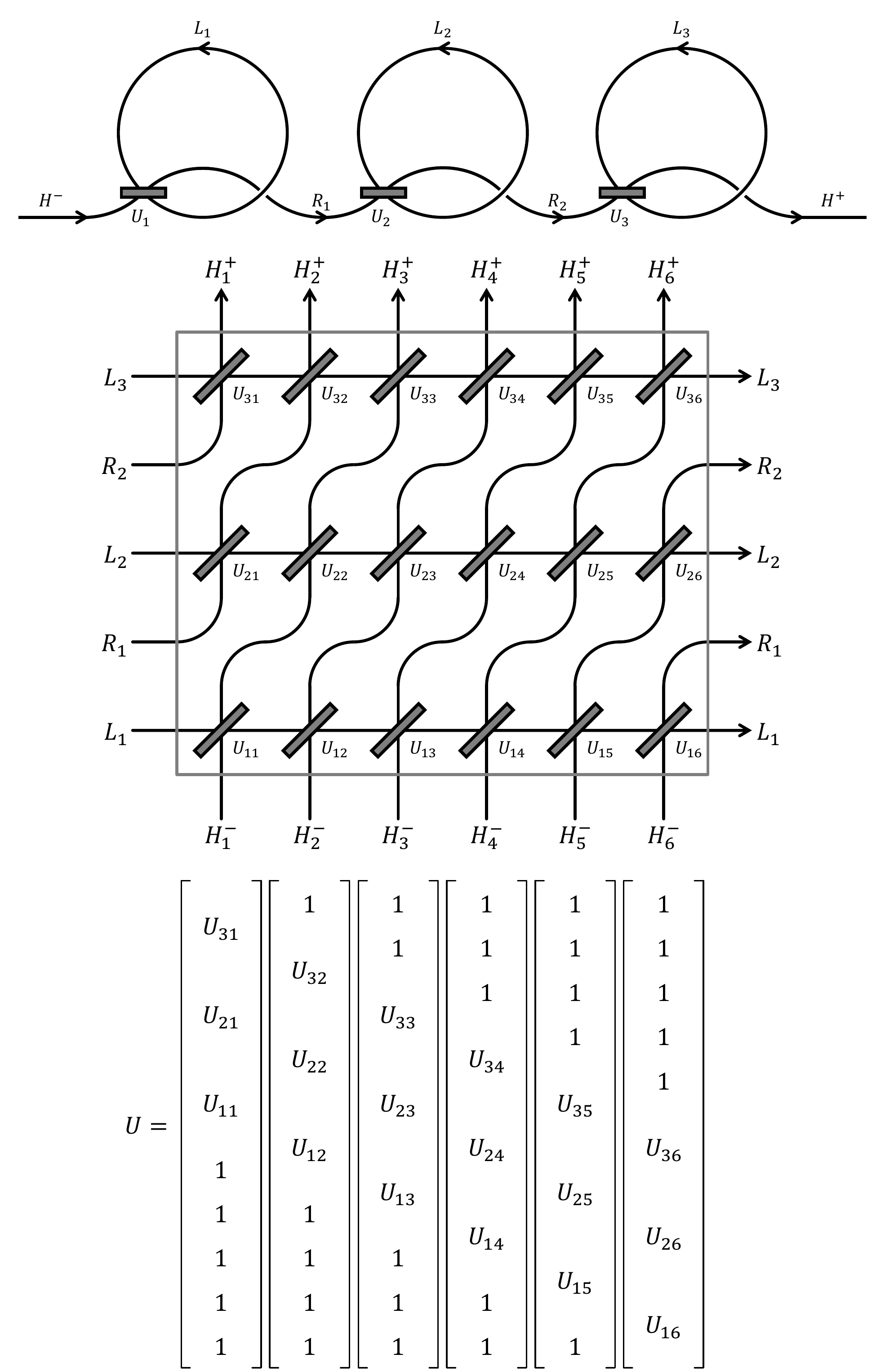}
\caption{The single-rail photonic computer. The physical circuit comprises a
sequence of $N_L$ loops ($N_L=3$ in this example) executed over $N_H$
consecutive time-bins ($N_H=6$ in this example). The generating unitary
matrix of the operation is the product of $N_H$ block-diagonal unitary
matrices, represented here as the column vectors of their diagonal blocks.}
\end{figure}

\paragraph{The single-rail photonic computer.}

The single-rail circuit is constructed by linking single-loop circuits at
their external edges. Executing the circuit over successive time-bins
generates a computer that is programmable through the configurations of the
beam splitters at each stage.

The single-rail photonic computer with $N_{L}$ loops has a total of $%
2N_{L}+1 $ edges: one edge each as input and output; $N_{L}$ edges forming
the loops that hold the internal state for computation; and a rail
comprising $N_{L}-1$ linking edges that enable writing onto and reading from
the loops. Each loop connects to the rail via a beam splitter. Executed over 
$N_{H}$ consecutive time-bins, the computer thus has $N_{L}N_{H}$
programmable beam splitters creating a total operation generated by a
unitary $(2N_{L}-1+N_{H})$-matrix. Since the operation in each beam splitter
is associated with a unitary $2$-matrix, the total operation is generated by
the product of $N_{H}$ block-diagonal unitary $(2N_{L}-1+N_{H})$-matrices,
with non-trivial blocks given by the generating matrices of the beam
splitters.

\begin{figure}[!t]
\centering\includegraphics[width=\linewidth]{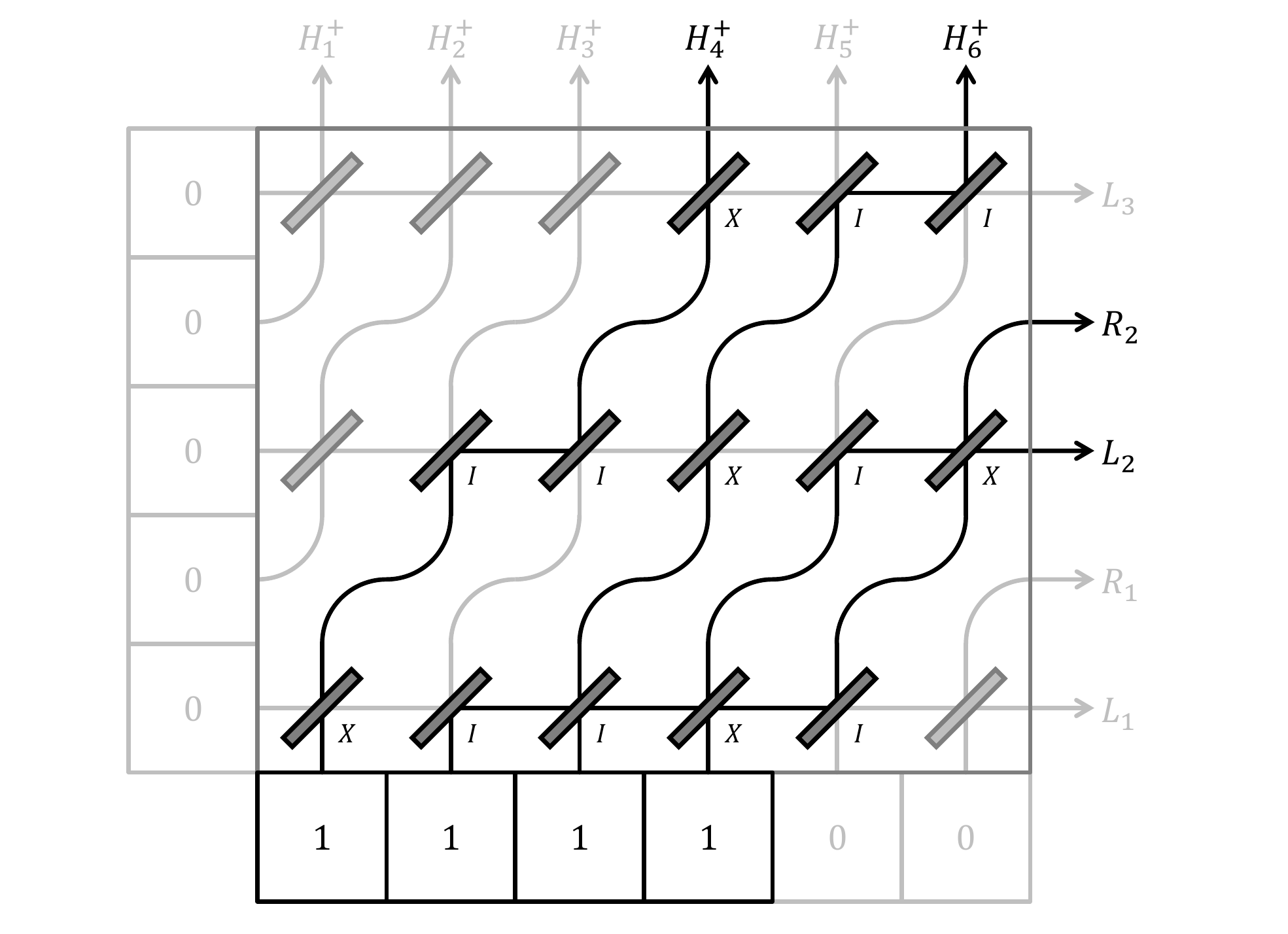}
\caption{Tracking the input photons through a deterministic circuit with
beam splitters configured either as mirrors or as windows, the resulting
operation permutes the output photons. More generally, beam splitters create
mixed states of photons on the output and internal edges in permutations
determined by the topology of the graph. Complexity of this operation
increases exponentially with the number of photons.}
\end{figure}

When the beam splitters are configured to be either mirrors or windows,
photons entered on the input edges have a deterministic path through the
circuit diagram. The map from input to output is then a permutation, mapping
each input edge either to a unique output edge or to an internal edge. Any
other configuration complicates this picture, as the beam splitter
simultaneously acts as mirror and as window, leaving the incident photon in
a mixed state.

Tracking the output from a defined input is an exercise in combinatorics,
bifurcating at each beam splitter to create an exponentially increasing set
of possible outcomes. With sufficient depth the circuit creates a
configurable map from input to output that is prohibitively expensive to
replicate on a classical computer. When used in a hybrid approach that
combines quantum computation with classical optimisation, this boson
sampling scheme can be incorporated into machine learning applications, and
is a leading candidate for noisy intermediate-scale quantum computing.

\section{Universal quantum computing}

Standard operations are abstracted in a theoretical model of quantum
computing that allows algorithms to be developed independently from the
physical computer. The theoretical model is implemented in the physical
model via operations that represent theoretical states as physical states
and interpret physical states as theoretical states, and a map that
implements theoretical operations as physical operations.

\begin{definition*}[Implementation]
The implementation of the theoretical model in the physical model associates
the theoretical system $Q$ with a physical system $[Q]$ via operations that
represent and interpret state:%
\begin{align}
\maprepresent& :Q\rightarrow [Q] \\
\mapinterpret& :[Q]\rightarrow Q  \notag
\end{align}%
satisfying the compatibility condition:%
\begin{equation}
\maprepresent\circ \mapinterpret=\iota
\end{equation}%
so that $\maprepresent$ is monomorphic and $\mapinterpret$ is epimorphic.
The implementation associates the theoretical operation $U:Q^{-}\rightarrow
Q^{+}$ with a physical operation:%
\begin{equation}
[U]:[Q^{-}]\rightarrow [Q^{+}]
\end{equation}%
satisfying the weak functoriality condition:%
\begin{equation}
\maprepresent\circ [U_{1}]\circ [U_{2}]\circ \mapinterpret=\maprepresent%
\circ [U_{1}\circ U_{2}]\circ \mapinterpret
\end{equation}%
for compatible operations. This enforces the consistency of circuit building
in the theoretical and physical systems.

The theoretical operation $U$ is implemented in the physical model as the
operation $\maprepresent\circ [U]\circ \mapinterpret$ constructed in three
steps: represent the input theoretical state as a physical state; act on
this with the physical operation; interpret the output physical state as a
theoretical state. This implementation is \emph{exact} when it satisfies the
weak naturality condition:%
\begin{equation}
\maprepresent\circ [U]\circ \mapinterpret=U
\end{equation}
\end{definition*}

\begin{figure}[!t]
\centering\includegraphics[width=\linewidth]{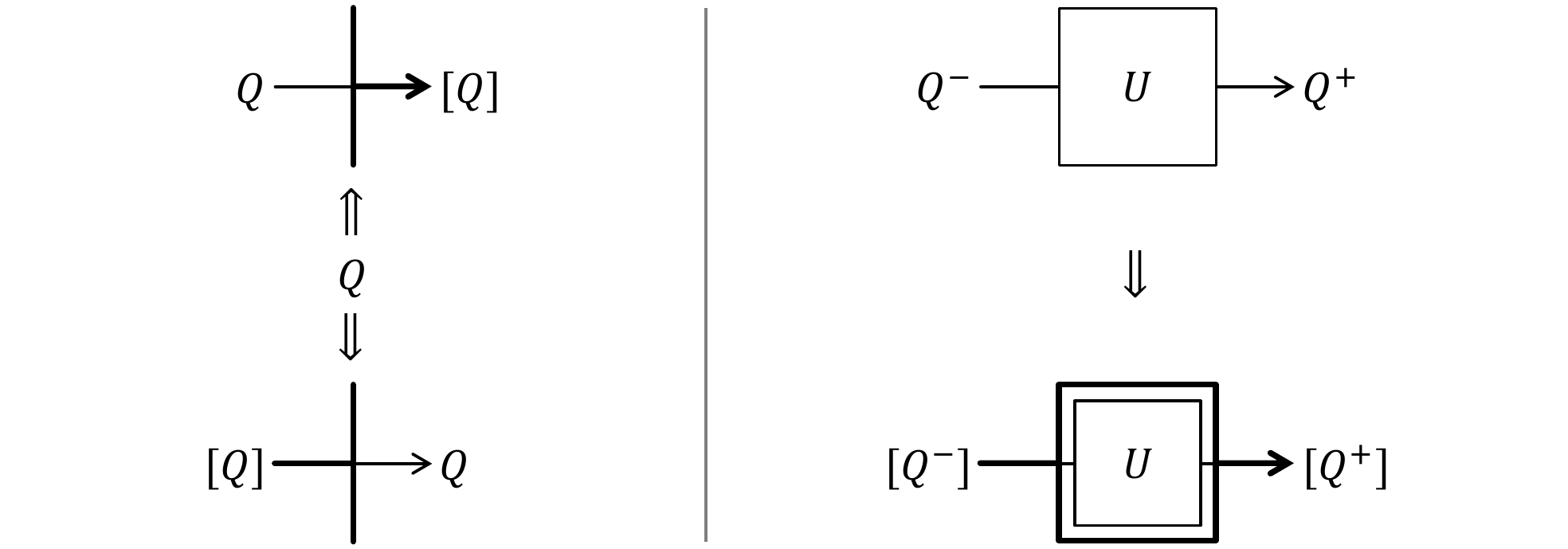}
\caption{Implementation of the theoretical model in the physical model. The
theoretical system $Q$ is located in the physical system $[Q]$ via an
operation that represents theoretical states as physical states and an
operation that interprets physical states as theoretical states. The
theoretical operation $U$ is then implemented as a physical operation $[U]$.
The implementation is exact when these constructions satisfy the coherence
properties of natural transformations and functors.}
\end{figure}

\begin{figure}[!t]
\centering\includegraphics[width=\linewidth]{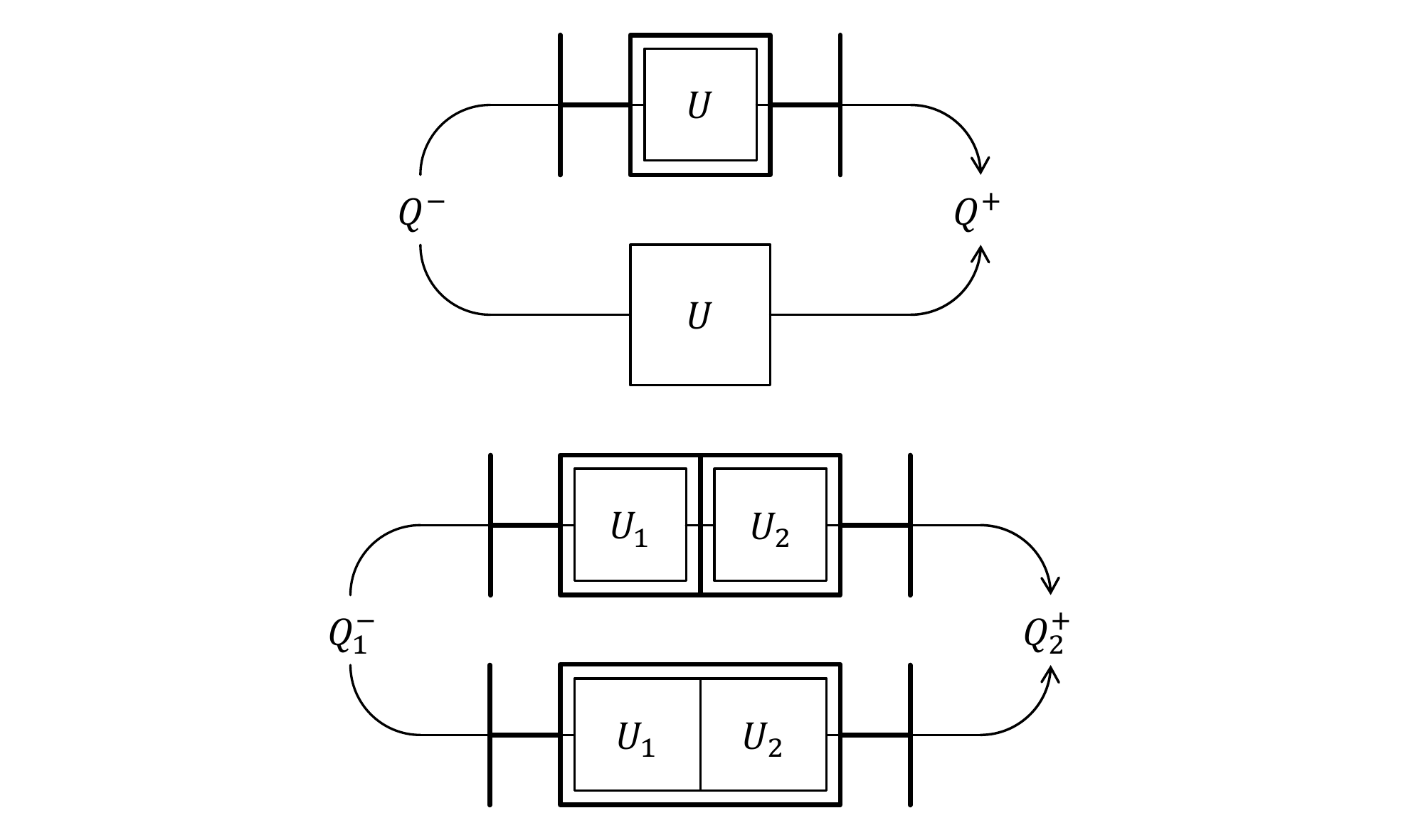}
\caption{The weak naturality and weak functoriality properties of the
implementation, expressed in these commutative diagrams, ensure that
algorithms developed in the theoretical model map precisely to the physical
model.}
\end{figure}

\begin{figure}[!p]
\begin{minipage}{\textwidth}
\begin{equation*}
\begin{array}{ccp{0.4\linewidth}}
\begin{tikzcd} & {[Q]} \arrow[rd,"{\smapinterpret}"] & \\ Q
\arrow[ru,"{\smaprepresent}"] \arrow[rr, "{\iota}"'] & & Q \end{tikzcd} & \hspace{%
1.4	cm} & \adjustbox{valign=c}{\includegraphics[width=%
\linewidth]{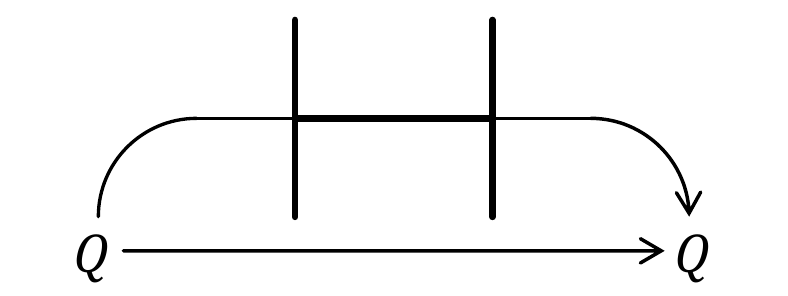}}%
\end{array}%
\end{equation*}%
\caption{The properties of an implementation are expressed as commutative diagrams, in traditional form on the left and in circuit form on the right. This diagram expresses compatibility of the operations that represent and interpret states. The diagrams below describe categorical properties that are sufficient to imply exactness for the implementation.}
\begin{equation*}
\begin{array}{ccp{0.45\linewidth}}
\begin{tikzcd} & {[Q^-]} \arrow[rd,"{[U]}"] & \\ Q^-
\arrow[ru,"{\smaprepresent}"] \arrow[rd,"{U}"'] & & {[Q^+]} \\ & Q^+ \arrow[ru,
"{\smaprepresent}"'] & \end{tikzcd} & \hspace{0.5cm} & \adjustbox{valign=c}{%
\includegraphics[width=\linewidth]{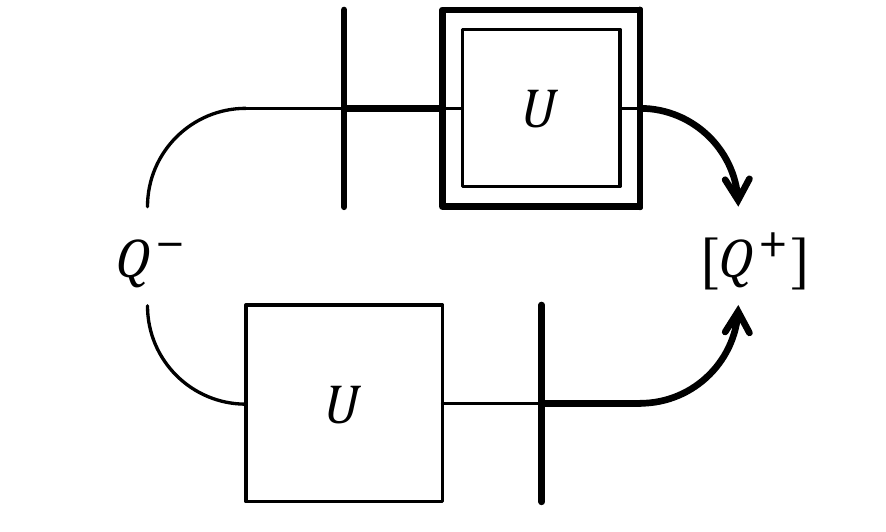}} \\ 
\begin{tikzcd} & Q^- \arrow[rd,"{U}"] & \\ {[Q^-]} \arrow[ru,"{\smapinterpret}"]
\arrow[rd,"{[U]}"'] & & Q^+ \\ & {[Q^+]} \arrow[ru, "{\smapinterpret}"'] &
\end{tikzcd} & \hspace{0.5cm} & \adjustbox{valign=c}{\includegraphics[width=%
\linewidth]{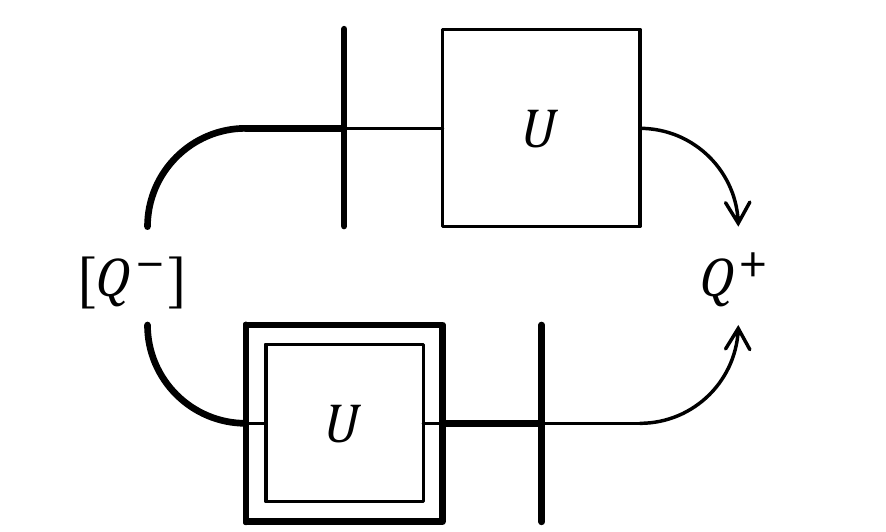}}%
\end{array}%
\end{equation*}%
\caption{These commutative diagrams express naturality of the operations that represent and interpret states, ensuring that an operation has the same effect on the state in the theoretical and physical models.}
\begin{equation*}
\begin{array}{ccp{0.45\linewidth}}
\begin{tikzcd}[column sep = small] & {[Q^+_1]=[Q^-_2]} \arrow[rd,"{[U_2]}"]
& \\ {[Q^-_1]} \arrow[ru,"{[U_1]}"] \arrow[rr, "{[U_1\circ U_2]}"'] & &
{[Q_2^+]} \end{tikzcd} & \hspace{0.1cm} & \adjustbox{valign=c}{%
\includegraphics[width=\linewidth]{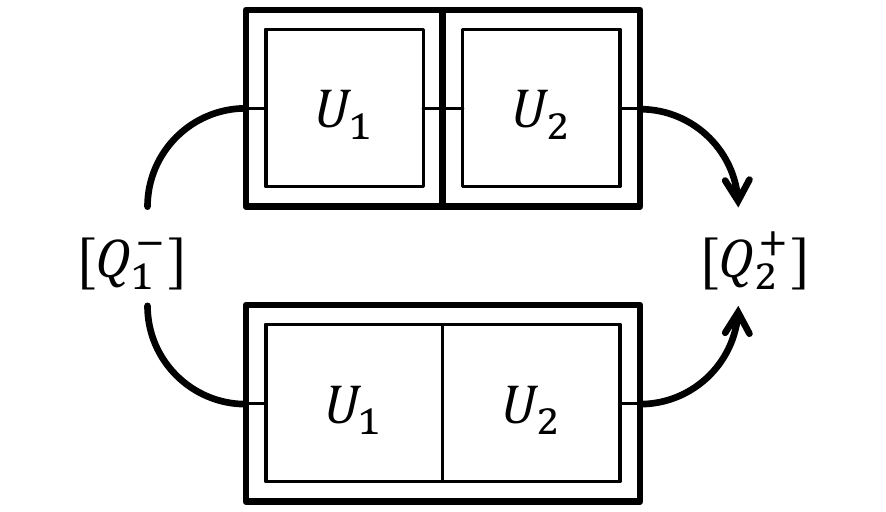}}%
\end{array}%
\end{equation*}%
\caption{This commutative diagram expresses functoriality of the implementation of operations, ensuring that circuit construction in the
theoretical model is mirrored in the physical model.}
\end{minipage}
\end{figure}

Taken together, weak functoriality and weak naturality ensure that
algorithms developed in the theoretical model map precisely to the physical
model. A sufficient (but not strictly necessary) condition is that the
implementation is a functor:%
\begin{equation}
[U_{1}\circ U_{2}]=[U_{1}]\circ [U_{2}]
\end{equation}%
and that the representation $\maprepresent$ and interpretation $%
\mapinterpret
$ are compatible natural transformations:%
\begin{align}
U\circ \maprepresent& =\maprepresent\circ [U] \\
\mapinterpret\circ U& =[U]\circ \mapinterpret  \notag
\end{align}%
Naturality implies that an operation has the same effect on the state
whether performed in the theoretical or physical model. Functoriality means
that circuit construction in the theoretical model is mirrored in the
physical model. These coherence properties are weakened by restricting to
the domain of application strictly required by the implementation.

For quantum systems, the representation ${\maprepresent}$ of theoretical
states as physical states is assumed to be an isometry:%
\begin{equation}
{\maprepresent\,\maprepresent}^{\ast }=\iota
\end{equation}%
The interpretation ${\mapinterpret}$ of physical states as theoretical
states is then taken to be the adjoint of the isometry:%
\begin{equation}
{\mapinterpret}={\maprepresent}^{\ast }
\end{equation}%
These maps are compatible thanks to the isometry property.

Exactness is a strong condition, and there are weaker conditions that can be
imposed on the quantum implementation while still partially maintaining its
utility. For the theoretical operation $U$ with physical implementation $[U]$%
, define the floor $f$ and cap $c$ to be respectively the largest and
smallest scalars satisfying the inequality:%
\begin{equation}
f\left\vert {\braket{b^-\maprepresent [U]\mapinterpret b^+}}\right\vert
^{2}\leq \left\vert {\braket{b^-|U|b^+}}\right\vert ^{2}\leq c\left\vert {%
\braket{b^-\maprepresent [U]\mapinterpret b^+}}\right\vert ^{2}
\end{equation}%
for all theoretical states ${\bra{b^-}}$ and ${\bra{b^+}}$, where ${%
\brarepresent{b}}$ denotes the physical state that represents the
theoretical state ${\bra{b}}$. The squared moduli of theoretical matrix
elements are in this way bounded by computable quantities. Noting that many
important quantum algorithms depend only on whether an outcome has non-zero
probability, the physical operation $[U]$ is said to be \emph{equivalent} to
the theoretical operation $U$ if these scalars satisfy:%
\begin{equation}
0<f\leq c<\infty
\end{equation}%
The physical computer estimates the matrix elements of $[U]$ via repeated
execution; equivalence then translates this into bounds for the matrix
elements of $U$ in the theoretical computer. Performance of the computer
thus depends on the accuracy of the physical estimate and the tightness of
the theoretical bounds.

Calculations on the physical computer are restricted to the squared moduli
of matrix elements in the computation basis of the system. It may not be
possible to precisely perform the physical calculation $\left\vert {%
\braket{b^-\maprepresent [U]\mapinterpret b^+}}\right\vert $, but
estimatable bounds can be identified:%
\begin{equation}
\mu -\varepsilon \leq \left\vert {\braket{b^-\maprepresent [U]\mapinterpret
b^+}}\right\vert ^{2}{\leq \mu +\varepsilon }
\end{equation}%
where:%
\begin{align}
\mu & =\sum_{\bra{n^-}}\sum_{\bra{n^+}}\left\vert {\braket{b^-\maprepresent
n^-}}\right\vert ^{2}\left\vert {\braket{n^-|[U]|n^+}}\right\vert
^{2}\left\vert {\braket{n^+\mapinterpret b^+}}\right\vert ^{2} \\
\mu +\varepsilon & =(\sum_{\bra{n^-}}\sum_{\bra{n^+}}\left\vert {%
\braket{b^-\maprepresent n^-}}\right\vert \left\vert {\braket{n^-|[U]|n^+}}%
\right\vert \left\vert {\braket{n^+\mapinterpret b^+}}\right\vert )^{2} 
\notag
\end{align}%
summing over the preparation and measurement bases of the physical computer.
Combined, the bounds on the theoretical calculation $\left\vert {%
\braket{b^-|U|b^+}}\right\vert $ are then:%
\begin{equation}
f(\mu -\varepsilon )\leq \left\vert {\braket{b^-|U|b^+}}\right\vert ^{2}\leq
c(\mu +\varepsilon )
\end{equation}%
Compatibility of the theoretical and physical computation bases thus
contributes to the overall performance of the quantum computer. In
particular, $\varepsilon =0$ when each theoretical basis state maps to an
unmixed physical basis state.

In the theoretical model of universal quantum computing the elementary
system is the qudit whose state space $Q$ is $d$-dimensional with
computation basis:%
\begin{equation}
\{\bra{b}:b=0,\ldots ,d-1\}
\end{equation}%
where $d$ is the arity of the model. Operations on this system are elements
of the group $\mathsf{U}[d]$, the unitary maps in $d$ dimensions. Encoding
these dits in the $d$-ary expansion of a positive integer, the state space
of the $N$-qudit system $\otimes ^{N}Q$ is $d^{N}$-dimensional with
computation basis:%
\begin{equation}
\{\bra{b}:b=0,\ldots ,d^{N}-1\}
\end{equation}%
Operations on this system are elements of the group $\mathsf{U}[d^{N}]$, the
unitary maps in $d^{N}$ dimensions. Determining whether a physical model is
able to represent these operations is simplified by reducing to a smaller
set of operations that generates them. Conveniently, operations of the qudit
model are generated by unitary maps on a single qudit:%
\begin{gather}
\includegraphics[width=\linewidth]{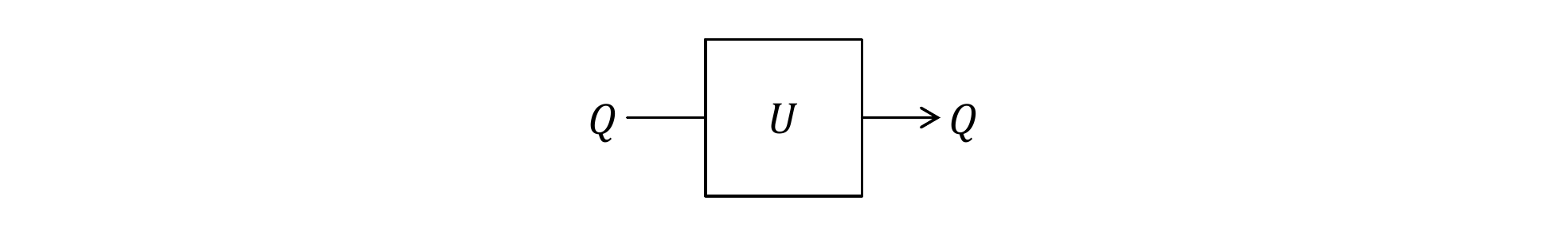}  \notag \\
U\in \mathsf{U}[d]
\end{gather}%
and any unitary map on two qudits that does not factorise as the
concatenation of two single-qudit maps. For the qubit ($d=2$) model, a
simple example is provided by the controlled-Z gate on two qubits:%
\begin{gather}
\includegraphics[width=\linewidth]{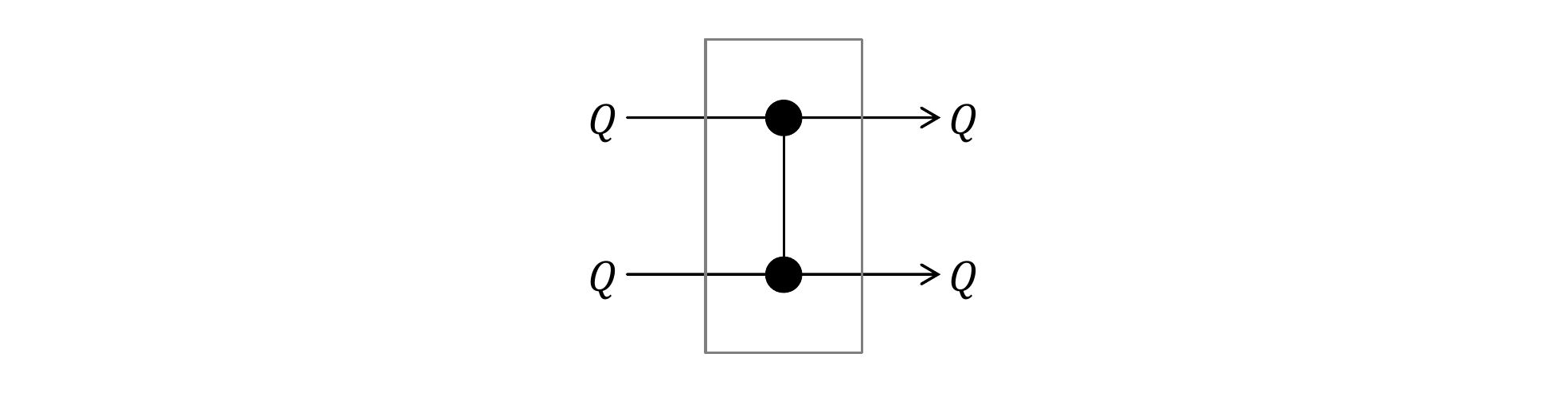}  \notag \\
C\!Z=%
\begin{bmatrix}
1 & 0 & 0 & 0 \\ 
0 & 1 & 0 & 0 \\ 
0 & 0 & 1 & 0 \\ 
0 & 0 & 0 & -1%
\end{bmatrix}%
\in \mathsf{U}[4]
\end{gather}%
Any physical model that implements this reduced set of operations implements
the universal model of quantum computing.

There are many physical systems that support this theoretical model, and
within a given physical system there may be multiple distinct ways to
implement the model, each with its advantages and disadvantages.
Specialising to photonic systems, two implementations are reviewed below:
the \emph{single-photon qubit} model, and the \emph{parity qudit} model.

\paragraph{The single-photon qubit model.}

In this model, the state space $Q$ of the single qubit is located within the
dual-mode photonic state space $[Q]=B\otimes B$. Theoretical states are
represented as photonic states via the isometry:%
\begin{align}
\brarepresent{0}& =\bra{01} \\
\brarepresent{1}& =\bra{10}  \notag
\end{align}%
and photonic states are interpreted as theoretical states via the adjoint of
this representation:%
\begin{align}
\brainterpret{01}& =\bra{0} \\
\brainterpret{10}& =\bra{1}  \notag
\end{align}%
with the other computation basis states mapping to zero. Each qubit basis
state is identified with a unique computation basis state in the photonic
state space. The squared modulus of the theoretical matrix element can thus
be derived from the corresponding physical matrix element estimated by the
photonic computer.

The single-photon qubit model is ideally suited to a machine that reliably
generates and detects small numbers of photons. Physical qubits are prepared
from the photon source and beam splitters are used to implement the
operations of the theoretical model.

\begin{figure}[!p]
\begin{minipage}{\textwidth}
\centering\includegraphics[width=\linewidth]{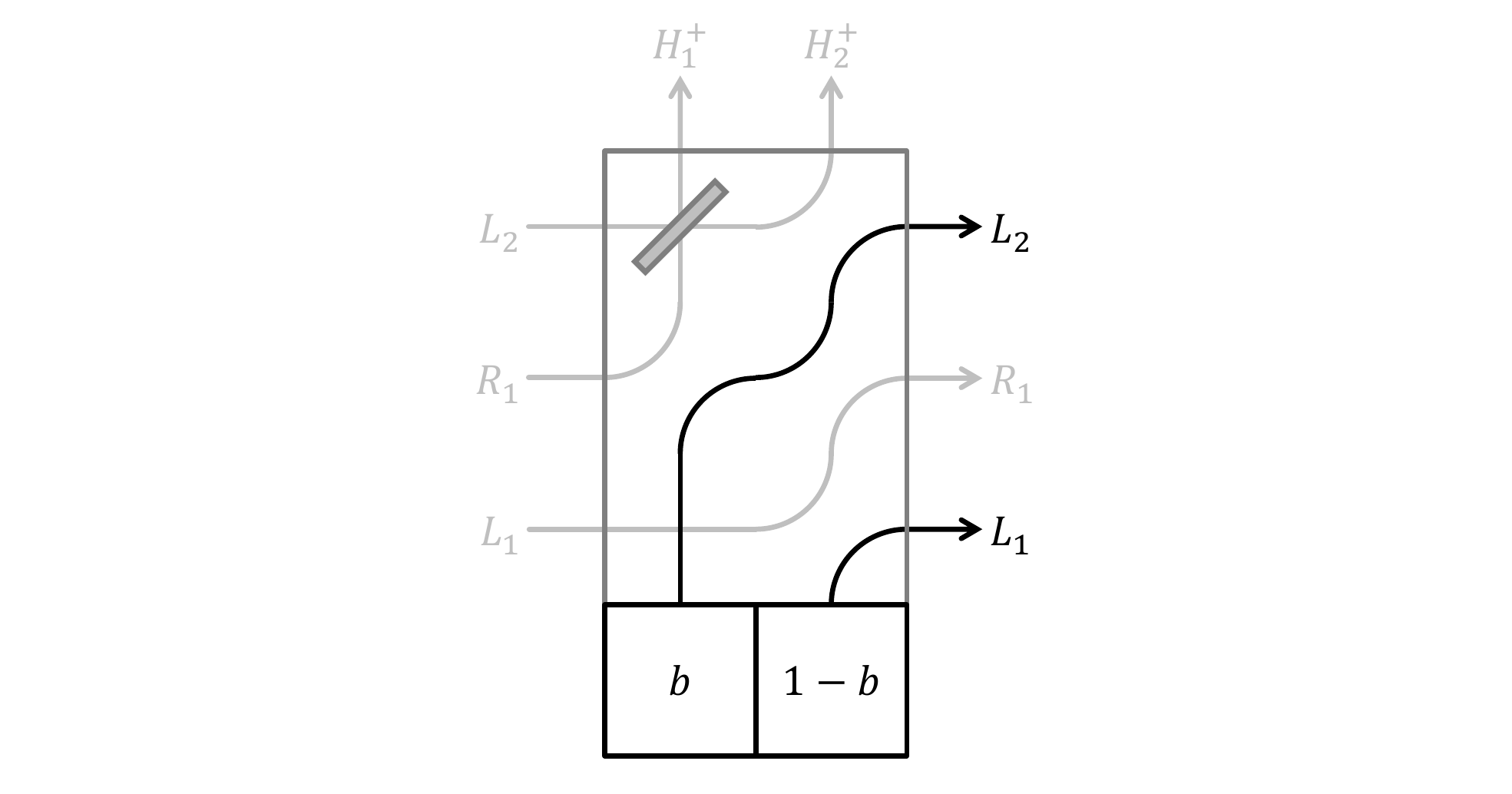}
\caption{The abstract constructions of the single-photon qubit model can be implemented on the single-rail photonic computer. The qubit basis state $\bra{b}$ is represented on two loops executed over
two time-bins.}
\end{minipage}
\newline
\newline
\newline
\begin{minipage}{\textwidth}
\centering\includegraphics[width=\linewidth]{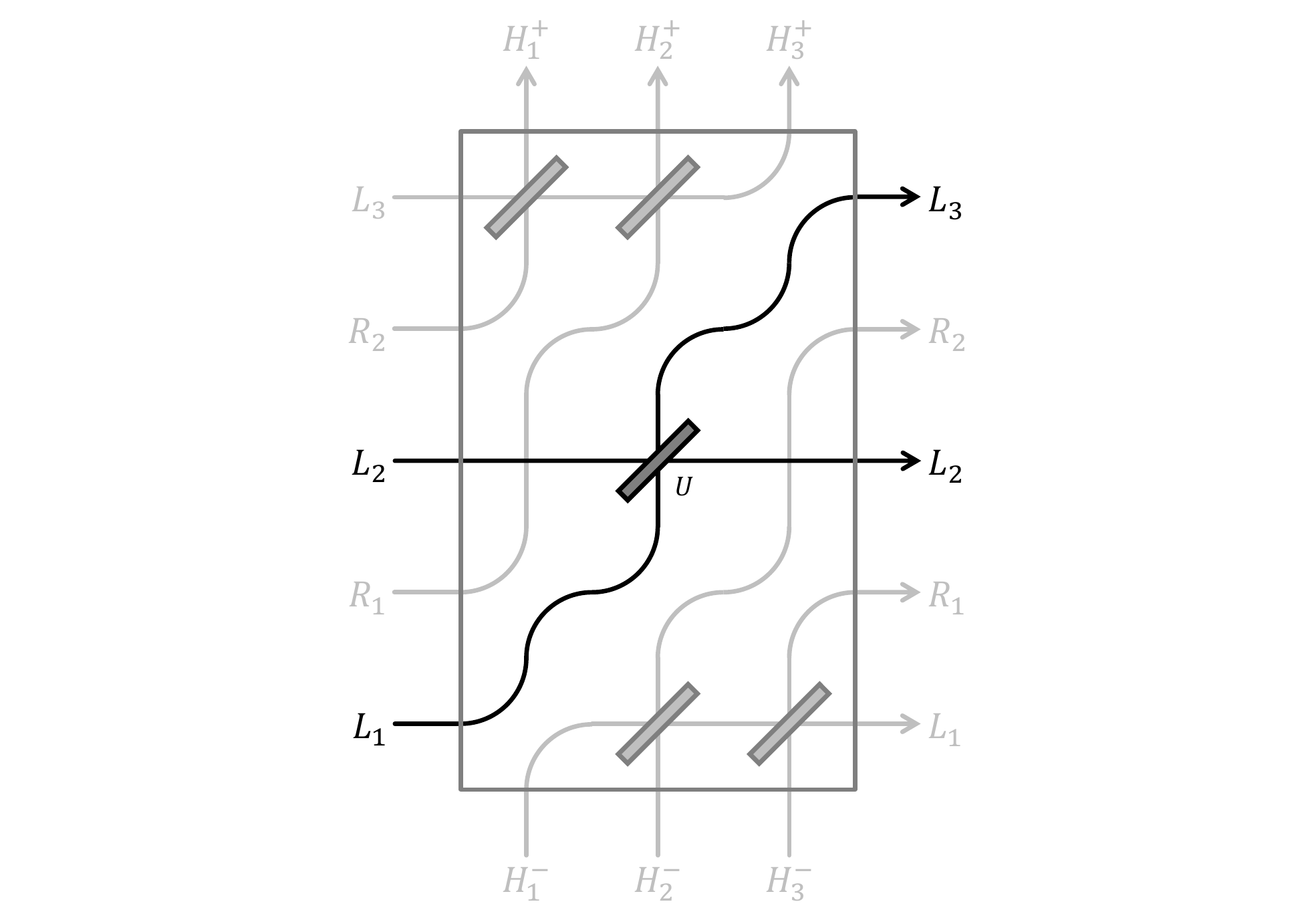}
\caption{The theoretical operation $U$ on a single qubit is implemented as a physical
operation on the single-rail photonic computer with three loops
executed over three time-bins. The input qubit is represented on the first
and second loops and the output qubit is interpreted from the second and
third loops. The beam splitter in the middle time-bin is configured to match
the target operation.}
\end{minipage}
\end{figure}

The diagrams below show how to represent qubits and implement single-qubit
gates on the photonic computer. In the first diagram, the qubit is
represented on two internal edges by using the input edges to introduce a
single photon. In the second diagram, the single-qubit operation is
performed on two internal edges without the need for interaction with the
external system. Both these circuits deterministically implement the desired
operation.%
\begin{equation}
\includegraphics[width=\linewidth]{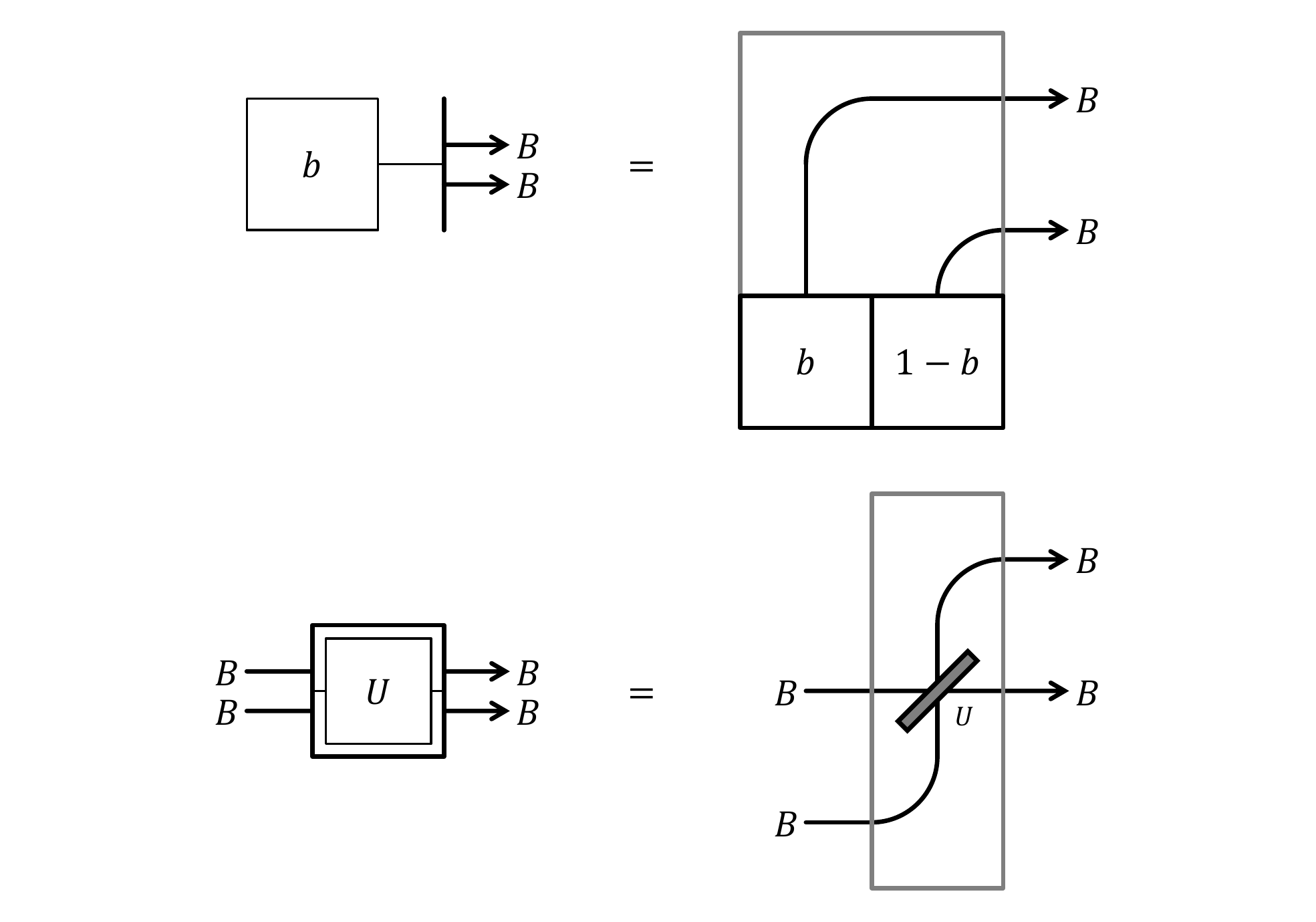}  \notag
\end{equation}%
Problems arise in this approach when implementing non-trivial operations on
multiple qubits, as the absence of photon interaction prevents these
operations from being implemented directly on the internal circuit. This
algorithmic shortfall is remediated by incorporating projective measurement,
at the cost of losing determinism in the outcome.

The protocol originally developed by Knill, Laflamme and Milburn illustrates
the key properties of quantum mechanics, and photonic systems in particular,
that contribute to the creation of multi-qubit operations. The scheme is
non-deterministic, requiring projective measurement, and makes use of
destructive interference and states with photon number greater than one to
implement the controlled-Z gate on two qubits.

There are two component operations that are combined in the scheme. The
first operation, the non-linear sign gate, uses projective measurement to
generate diagonal transformations on the basis states of a single mode. The
second operation, the Hadamard gate, uses destructive interference to
distinguish states in two modes, effectively using states with higher photon
count as holding places so that non-trivial diagonal transformations can be
applied to the two-qubit state. Composing these operations allows different
transformations to be applied to the four possible states of the two-qubit
system.

The non-linear diagonal gate:%
\begin{gather}
\includegraphics[width=\linewidth]{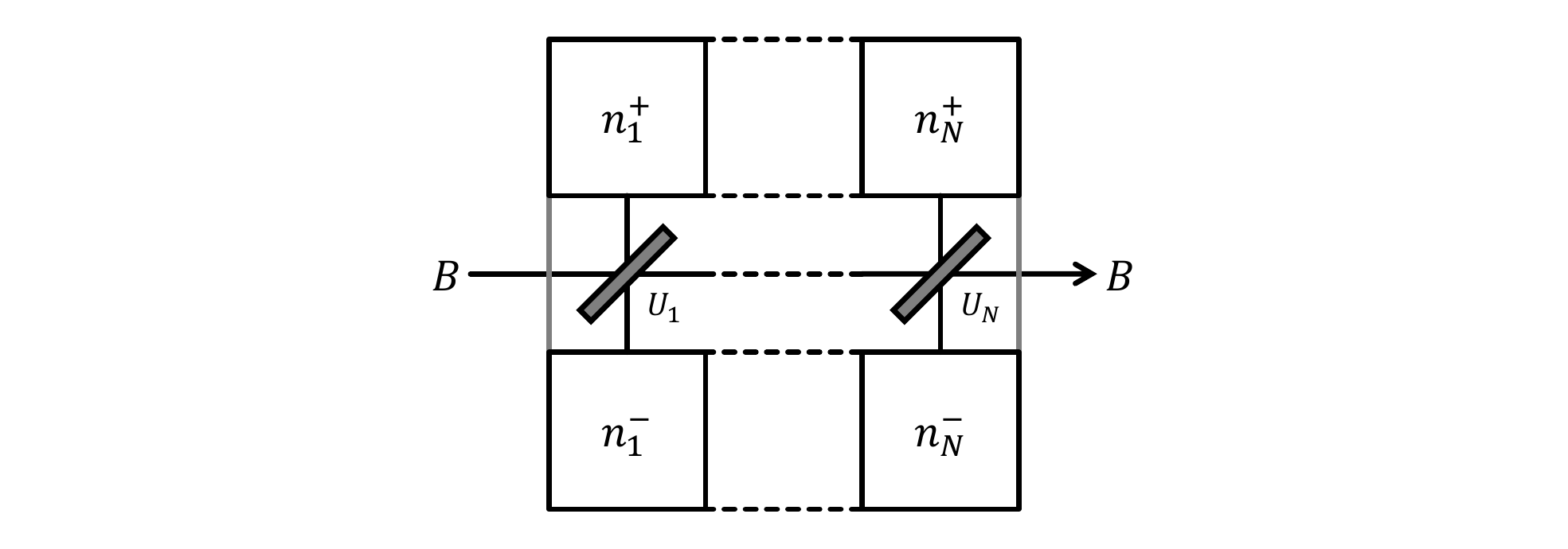}  \notag \\
N\!D=\braket{n^-_1|B[U_1]|n^+_1}\cdots \braket{n^-_N|B[U_N]|n^+_N}
\end{gather}%
projects the operation of temporally-composed beam splitters, and can be
implemented on the single-loop photonic computer. It is a non-deterministic
gate, depending on the photon counts on the output system, though the
probability of success can be optimised by using the feed-forward protocol.
When the total input and output photon counts match:%
\begin{equation}
n_{1}^{-}+\cdots +n_{N}^{-}=n_{1}^{+}+\cdots +n_{N}^{+}
\end{equation}%
conservation of photon number implies this gate is diagonal:%
\begin{equation}
\bra{m}\!N\!D=p[m]\bra{m}
\end{equation}%
The phases and angles of the unitary $2$-matrices associated with the beam
splitters are then fine tuned to the desired action on the basis states. As
a rule, each beam splitter introduces an angle that can be calibrated to a
single relationship among the diagonal coefficients.

Even though the number of internal photons is unchanged by the non-linear
diagonal gate, funneling the photons through a series of beam splitters
forces an interaction with the external system that effects a transformation
on the internal state. Aside from elementary cases, it is not possible to
create these operations using linear optical components without projective
measurement.

An important example is the non-linear sign gate $N\!S$, engineered using
two beam splitters and one ancillary photon to impose the following
relationships on states with low photon number:%
\begin{align}
\bra{0}\!N\!S& =p\bra{0} \\
\bra{1}\!N\!S& =p\bra{1}  \notag \\
\bra{2}\!N\!S& =-p\bra{2}  \notag
\end{align}%
for a positive scalar $p$. This operation is closed on states with at most
two photons, implementing a sign change on the state with two photons. It is
necessarily non-deterministic, generating the target operation on states in
this subspace with probability $p^{2}$. Imposing two relationships between
the coefficients, the gate is implemented as a non-linear diagonal gate
using two temporally-composed beam splitters in a circuit of the form:%
\begin{gather}
\includegraphics[width=\linewidth]{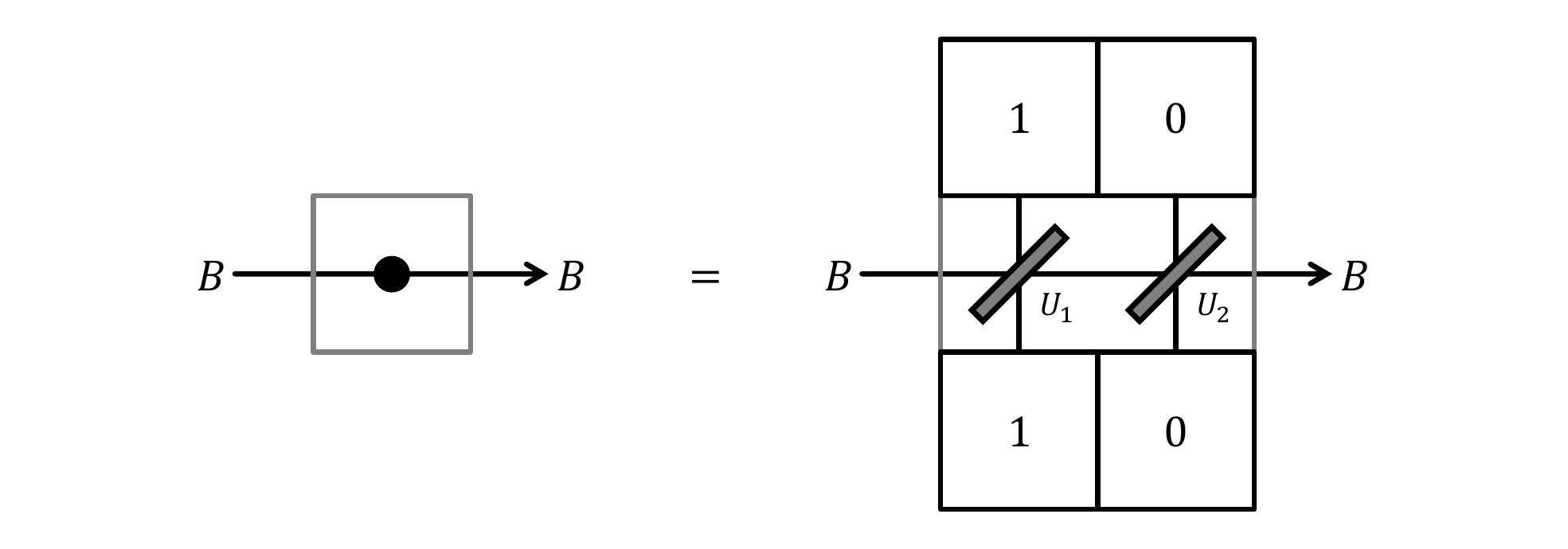}  \notag \\
N\!S=\braket{1|B[U_1]|1}\braket{0|B[U_2]|0}
\end{gather}%
To identify the unitary $2$-matrices satisfying the target diagonal
properties, first evaluate the operation for low photon number:%
\begin{align}
\bra{0}\!N\!S& =e^{i(\gamma _{1}+\tau _{1})}\cos [\theta _{1}]\bra{0} \\
\bra{1}\!N\!S& =-e^{i(2\gamma _{1}+\gamma _{2}-\tau _{2})}(1-2\cos [\theta
_{1}]^{2})\cos [\theta _{2}]\bra{1}  \notag \\
\bra{2}\!N\!S& =-e^{i(3\gamma _{1}-\tau _{1}+2\gamma _{2}-2\tau _{2})}\cos
[\theta _{1}](2-3\cos [\theta _{1}]^{2})\cos [\theta _{2}]^{2}\bra{2}  \notag
\end{align}%
then impose the phase and angle conditions:%
\begin{gather}
\gamma _{1}+\tau _{1}=2\gamma _{1}+\gamma _{2}-\tau _{2}=3\gamma _{1}-\tau
_{1}+2\gamma _{2}-2\tau _{2}=0 \\
\cos [\theta _{1}]=-(1-2\cos [\theta _{1}]^{2})\cos [\theta _{2}]=\cos
[\theta _{1}](2-3\cos [\theta _{1}]^{2})\cos [\theta _{2}]^{2}  \notag
\end{gather}%
The phase conditions are solved by:%
\begin{align}
\gamma _{1}& =-\tau _{1} \\
\gamma _{2}& =2\tau _{1}+\tau _{2}  \notag
\end{align}%
and the angle conditions are solved by:%
\begin{align}
\cos [\theta _{1}]& =\sqrt{\frac{3-\sqrt{2}}{7}} \\
\cos [\theta _{2}]& =-\sqrt{5-3\sqrt{2}}  \notag
\end{align}%
The non-linear sign gate then generates the target operation with
probability:%
\begin{equation}
p^{2}=\frac{3-\sqrt{2}}{7}
\end{equation}%
This construction makes use of both the core assumptions of quantum
mechanics, fine tuning the reflectance of the beam splitters to exploit the
interference of photons, and applying the projective effect on the internal
system of measurement on the external system.

The second operation contributing to the implementation of the controlled-Z
gate is the Hadamard single-qubit gate, physically constructed by a beam
splitter configured with generating unitary matrix that maximally mixes the
photon states:%
\begin{gather}
\includegraphics[width=\linewidth]{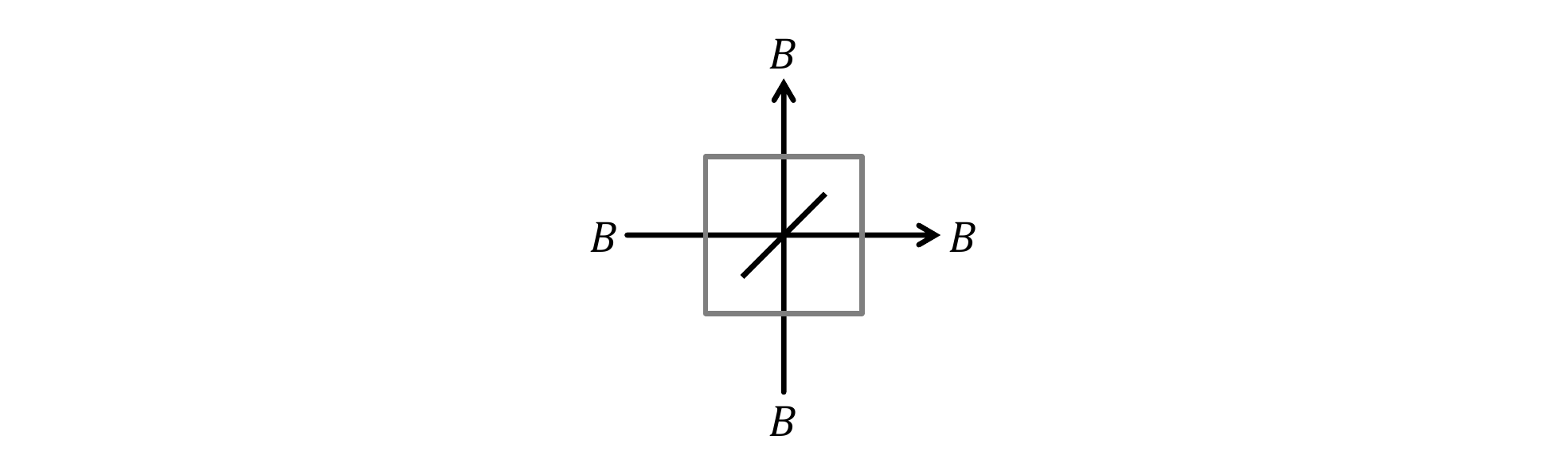}  \notag \\
H=\frac{1}{\sqrt{2}}%
\begin{bmatrix}
1 & 1 \\ 
1 & -1%
\end{bmatrix}%
\end{gather}%
This is an involutive operation whose action on states with low photon
number is given by:%
\begin{align}
\bra{00}\!B[H]& =\bra{00} \\
\bra{01}\!B[H]& =\frac{1}{\sqrt{2}}(\bra{10}-\bra{01})  \notag \\
\bra{10}\!B[H]& =\frac{1}{\sqrt{2}}(\bra{10}+\bra{01})  \notag \\
\bra{11}\!B[H]& =\frac{1}{\sqrt{2}}(\bra{20}-\bra{02})  \notag
\end{align}%
Destructive interference eliminates the state $\bra{11}$ in the fourth
expression, a property of photonic systems first demonstrated experimentally
by Hong, Ou and Mandel.

\begin{figure}[!p]
\centering\includegraphics[width=\linewidth]{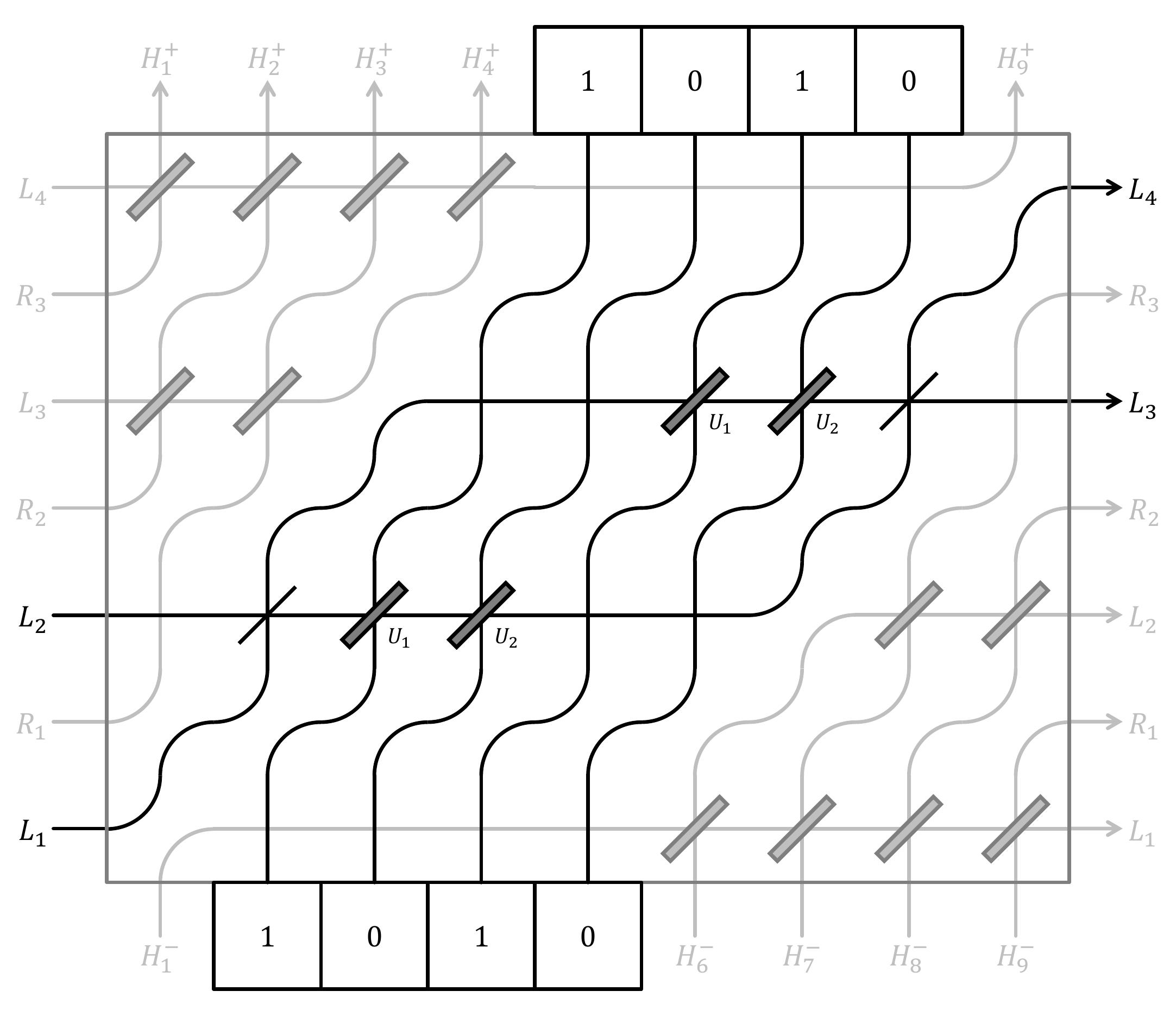}
\caption{The photonic controlled-Z operation $C\!Z$ on two qubits is
implemented as a physical operation on the single-rail photonic computer
with four loops executed over nine time-bins. The input qubits are
represented on the first and second loops and the output qubits are
interpreted from the third and fourth loops. Entanglement of the photons is
created by combining Hadamard and non-linear sign gates to flip the sign
when the initial internal state is $\bra{11}$. The scheme uses projective
measurement, and is only successful when the output photons are counted in
the specified amounts.}
\end{figure}

Non-linear sign gates are combined with Hadamard gates to create a
non-deterministic photonic version of the controlled-Z gate:%
\begin{gather}
\includegraphics[width=\linewidth]{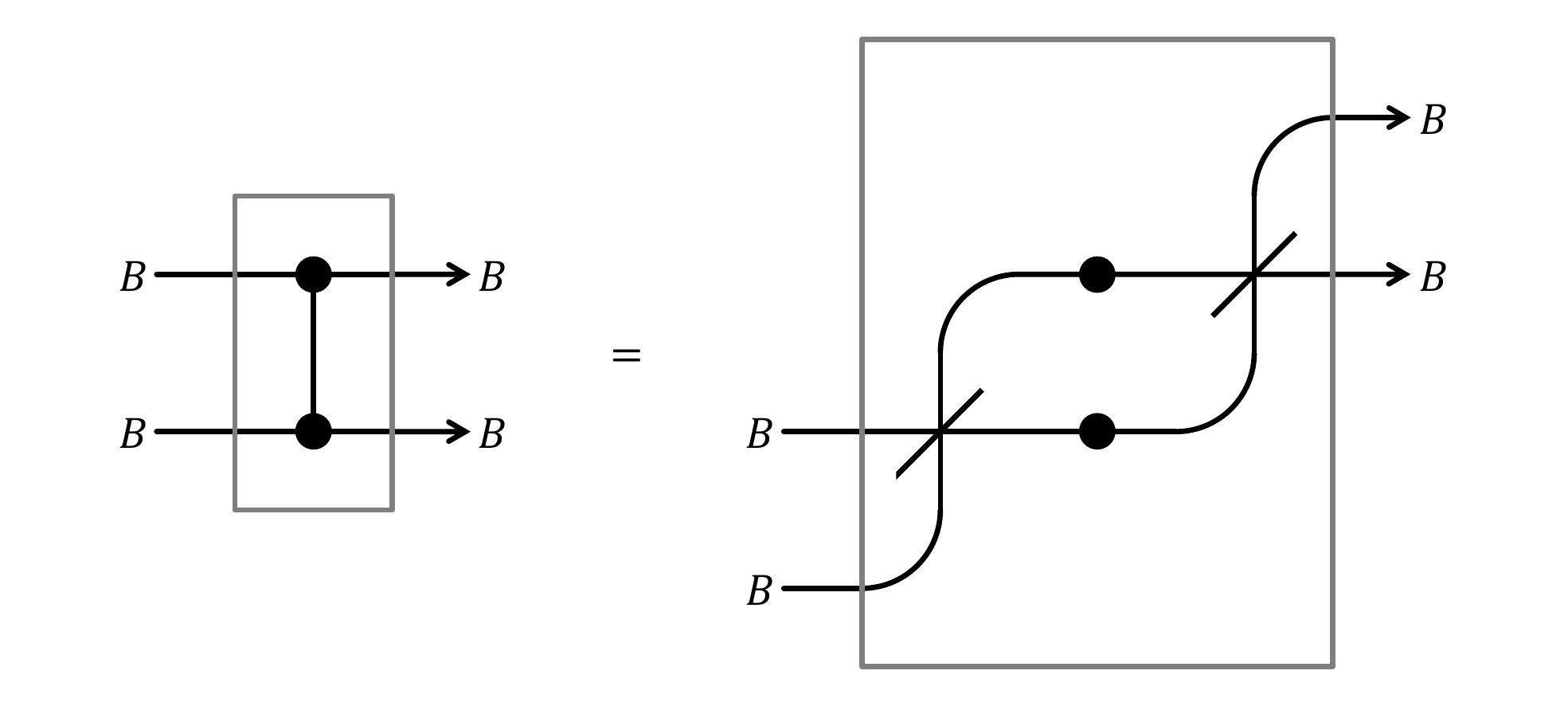}  \notag \\
C\!Z=B[H](N\!S\otimes N\!S)B[H]
\end{gather}%
The circuit contains two instances of the non-linear sign gate, generating
the target operation with probability:%
\begin{equation}
p^{4}=\frac{11-6\sqrt{2}}{49}
\end{equation}%
Tracking photons through the circuit, the photonic controlled-Z gate
switches the sign of the internal basis state $\bra{11}$ but leaves the
internal basis states $\bra{00}$, $\bra{01}$ and $\bra{10}$ unchanged. This
gate is integrated into a circuit that exactly implements the theoretical
controlled-Z gate:%
\begin{equation}
\includegraphics[width=\linewidth]{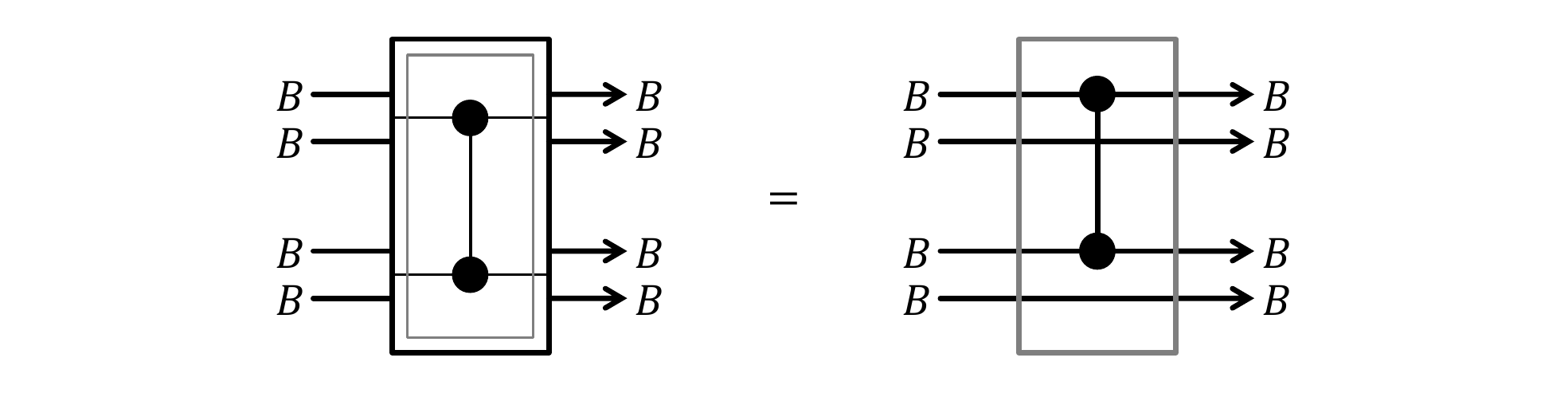}  \notag
\end{equation}

There are more efficient implementations of the controlled-Z gate in the
single-photon qubit model, further optimising the probability of success and
constructed on simpler photonic circuits. The commonalities among these
implementations are the fine tuning of the reflectances of beam splitters to
exploit interference and the application of projective measurement to
compensate for the lack of photon interaction. Success, which cannot be
guaranteed in the scheme, is then indicated by measurements of the ancillary
photons.

\paragraph{The parity qudit model.}

In this model, the state space $Q$ of the single qudit with arity $d$ is
located within the single-mode photonic state space $[Q]=B$. Theoretical
states are represented as photonic states via the isometry:%
\begin{equation}
\brarepresent{b}=\sum_{n\func{mod}d=b}\alpha _{n}\bra{n}
\end{equation}%
and photonic states are interpreted as theoretical states via the adjoint of
this representation:%
\begin{equation}
\brainterpret{n}=\alpha _{n}^{\ast }\bra{n\func{mod}d}
\end{equation}%
The model is parametrised by the sequence $(\alpha _{n})$ of scalars,
normalised to ensure that the representation is isometric:%
\begin{equation}
\sum_{n\func{mod}d=b}|\alpha _{n}|^{2}=1
\end{equation}%
With this representation of theoretical states as physical states, the
physical implementation $[U]$ of the theoretical operation $U$ on a single
qudit satisfies:%
\begin{equation}
\mu -\varepsilon \leq \left\vert {\braket{b^-\maprepresent [U]\mapinterpret
b^+}}\right\vert ^{2}{\leq \mu +\varepsilon }
\end{equation}%
where:%
\begin{align}
\mu & =\sum_{n^{-}\func{mod}d=b^{-}}\sum_{n^{+}\func{mod}d=b^{+}}|\alpha
_{n^{-}}|^{2}|\alpha _{n^{+}}|^{2}\left\vert {\braket{n^-|[U]|n^+}}%
\right\vert ^{2} \\
\mu +\varepsilon & =(\sum_{n^{-}\func{mod}d=b^{-}}\sum_{n^{+}\func{mod}%
d=b^{+}}|\alpha _{n^{-}}||\alpha _{n^{+}}|\left\vert {\braket{n^-|[U]|n^+}}%
\right\vert )^{2}  \notag
\end{align}%
Bounds for the squared modulus of the theoretical matrix element can thus be
derived from the corresponding physical matrix elements estimated by the
photonic computer.

Creating physical states that represent theoretical states is challenging in
this model, since they are mixtures of basis states that include all
possible photon counts. Fortunately there is a version of the model that has
a relatively straightforward physical implementation. This model is
specified by the representation and interpretation:%
\begin{align}
\brarepresent{b}& =\frac{1}{\sqrt{\func{exph}_{b}^{d}[|\alpha |^{2}]}}\sum_{n%
\func{mod}d=b}\frac{\alpha ^{n}}{\sqrt{n!}}\bra{n} \\
\brainterpret{n}& =\frac{1}{\sqrt{\func{exph}_{n\func{mod}d}^{d}[|\alpha
|^{2}]}}\frac{\alpha ^{\ast n}}{\sqrt{n!}}\bra{n\func{mod}d}  \notag
\end{align}%
parametrised by the non-zero scalar $\alpha $. The fractional exponential
functions in these expressions are defined by:%
\begin{equation}
\func{exph}_{b}^{d}[\theta ]:=\sum_{n\func{mod}d=b}\frac{\theta ^{n}}{n!}=%
\frac{1}{d}\sum_{c=0}^{d-1}\exp [e^{2\pi i(c/d)}\theta -2\pi i(bc/d)]
\end{equation}%
which reduce to the familiar hyperbolic functions in the binary case:%
\begin{align}
\func{exph}_{0}^{2}[\theta ]& =\cosh [\theta ] \\
\func{exph}_{1}^{2}[\theta ]& =\sinh [\theta ]  \notag
\end{align}

The parameter $\alpha $ is identified with the complex amplitude of a
canonical coherent state that represents a mixed theoretical state:%
\begin{gather}
\includegraphics[width=\linewidth]{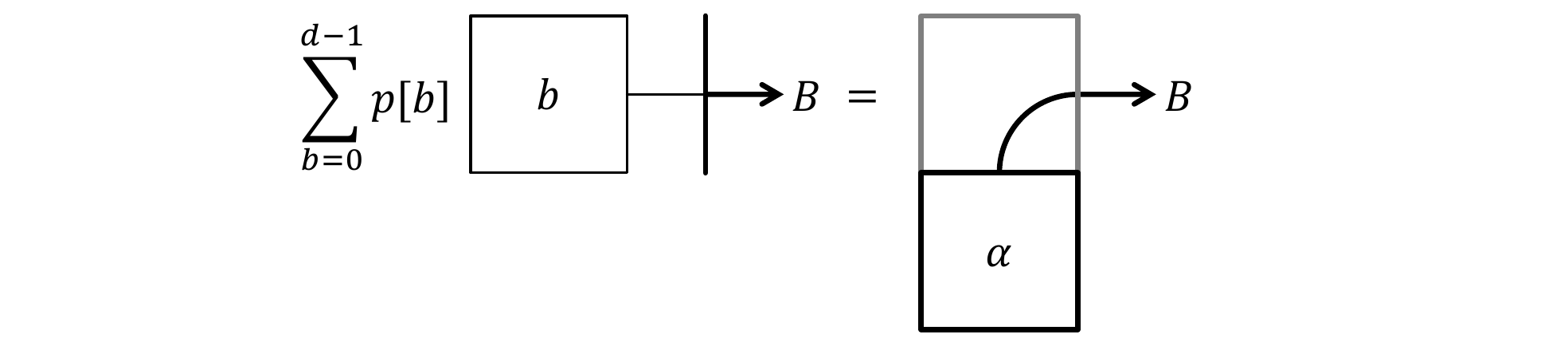}  \notag \\
\sum_{b=0}^{d-1}p[b]\brarepresent{b}=\bra{\alpha}
\end{gather}%
The mixing weights are:%
\begin{equation}
p[b]=\sqrt{\frac{\func{exph}_{b}^{d}[|\alpha |^{2}]}{\exp [|\alpha |^{2}]}}
\end{equation}%
and the canonical coherent state is expressed in terms of basis states as:%
\begin{equation*}
\bra{\alpha}=e^{-\frac{1}{2}|\alpha |^{2}}\sum_{n=0}^{\infty }\frac{\alpha
^{n}}{\sqrt{n!}}\bra{n}
\end{equation*}%
This is the physical state generated by an idealised laser. While it is
difficult to isolate the representation of an unmixed theoretical basis
state, it is possible to physically represent a mixed theoretical basis
state by calibrating the laser to the specified amplitude. The limiting
cases for this mixed theoretical state are:%
\begin{alignat}{2}
\bra{0}& & \hspace{0.5cm}& \text{(as }|\alpha |\rightarrow 0\text{)} \\
\frac{1}{\sqrt{d}}\sum_{b=0}^{d-1}\bra{b}& & \hspace{0.5cm}& \text{(as }%
|\alpha |\rightarrow \infty \text{)}  \notag
\end{alignat}%
The choice of parameter $\alpha $ for the model thus determines the mixing
of the theoretical state represented by the corresponding canonical coherent
state, where the state is unmixed for small $|\alpha |$ and fully mixed for
large $|\alpha |$.

\begin{figure}[!t]
\centering\includegraphics[width=\linewidth]{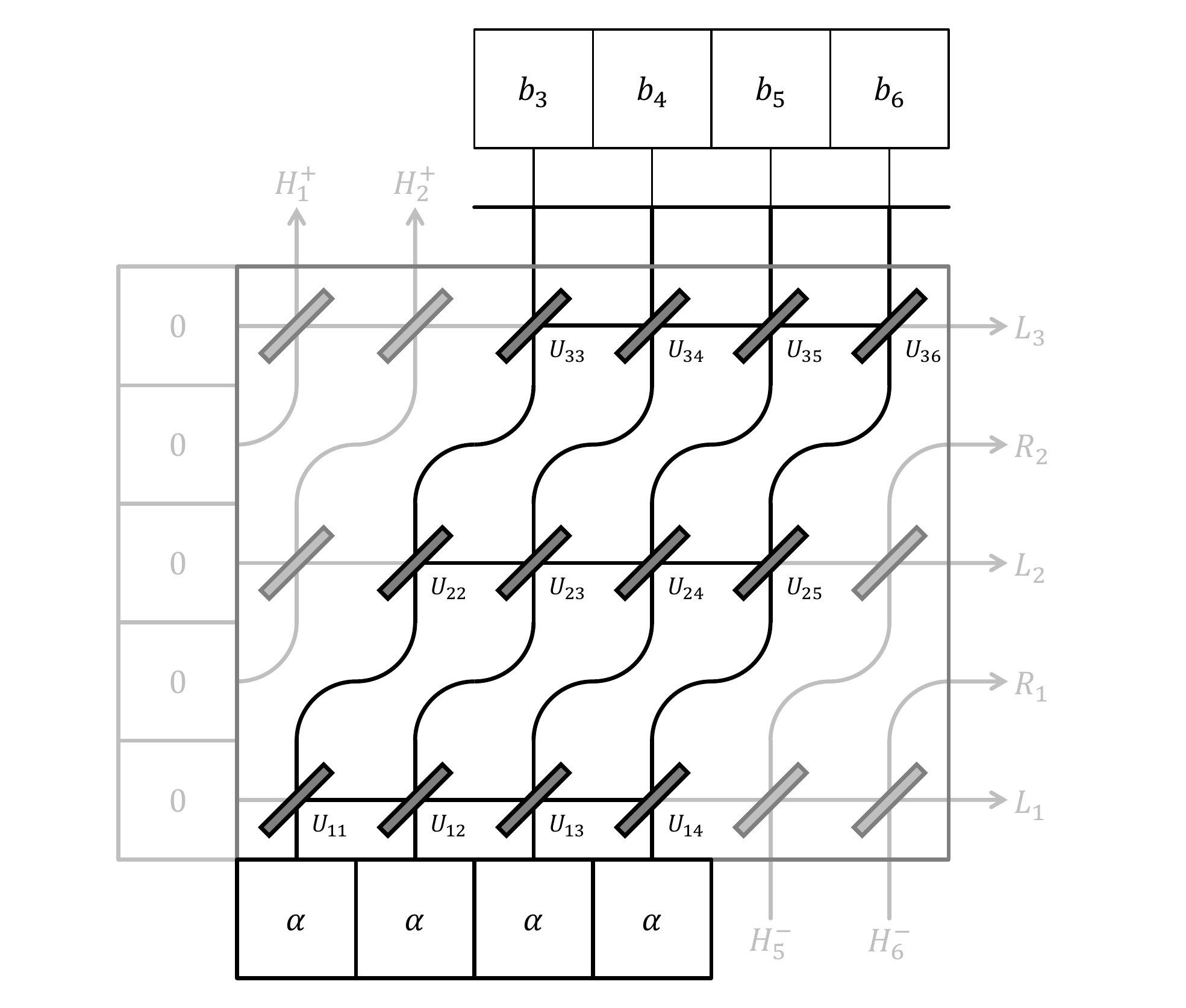}
\caption{A boson sampling scheme based on the parity qudit model,
implemented on the single-rail photonic computer with three loops executed
over six time-bins. The computer is seeded using an idealised laser on the
input edge and generates a ditstream on the output edge. In the application
to machine learning, the reflectances of the beam splitters encode data and
provide the free parameters used to train the computer.}
\end{figure}

Since the theoretical qudit is measured as the parity of the photon count,
efficacy of the quantum computer implemented by the parity qudit model
depends on the fidelity of the photon detector. A balance needs to be struck
between using an amplitude large enough to generate workable mixed states
while keeping the total photon count throughout the calculation low enough
to ensure its parity is distinguishable by the detector.

The many-to-one interpretation of physical basis states as theoretical basis
states means that theoretical matrix elements are only approximated by
physical matrix elements. In spite of this drawback, the model has the
advantage that any circuit can, in principle, be used as an implementation
of a theoretical operation, without requiring the fine tuning that enforces
only one photon on each physical qubit. Finding a family of such operations
that satisfies the coherence requirements of the implementation is far from
trivial, though this is less of an issue if the operations are used in
isolation, as is the case in boson sampling applications.

The beam splitter generates operations on one and two qudits:%
\begin{gather}
\includegraphics[width=\linewidth]{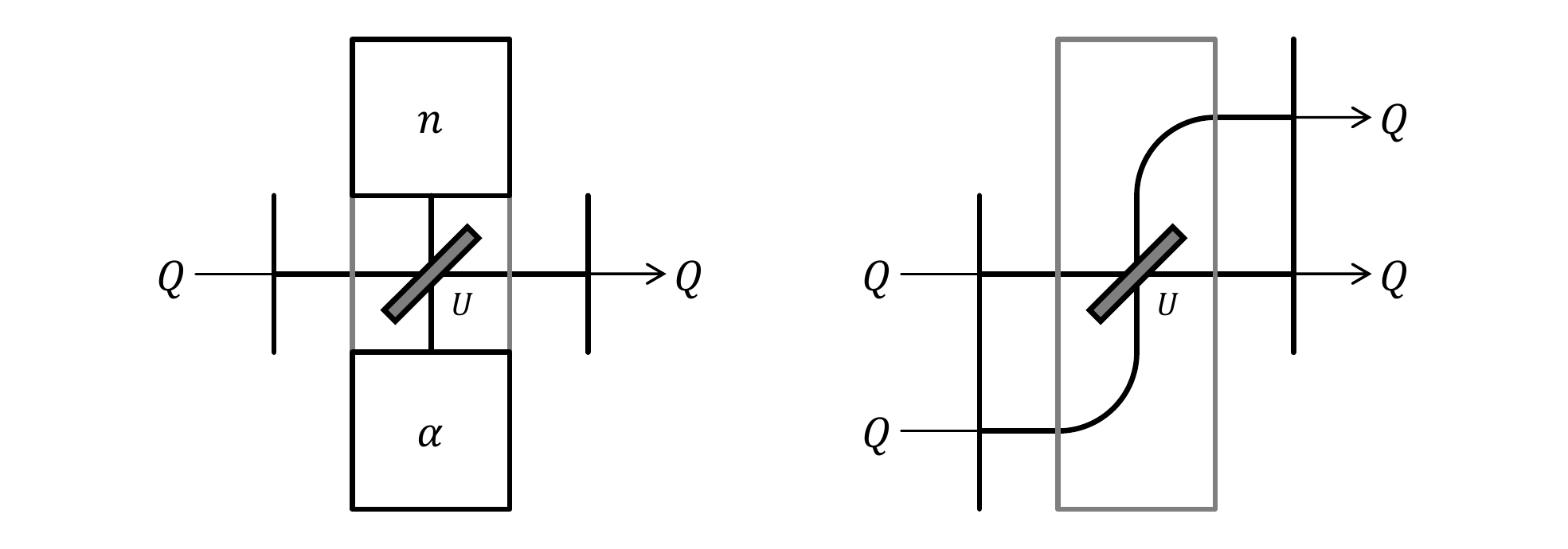}  \notag \\
\hspace{0.4cm}\maprepresent\braket{\alpha|B[U]|n}\mapinterpret\hspace{3cm}(%
\maprepresent\otimes \maprepresent)B[U](\mapinterpret\otimes \mapinterpret)
\end{gather}%
whose theoretical matrix elements are derived from the physical matrix
elements of the beam splitter. These are valid operations on the qudit state
space for any input and output photon counts on the physical circuit.
Photonic circuits constructed from beam splitters access the combinatorial
complexity of the full photonic state space, which grows exponentially as
the circuit is expanded. This creates a rich family of operations for the
parity qudit model.

\section{Conclusion}

The model of computing outlined in this essay uses photonic circuits to
perform calculations that are computationally expensive for classical
computers. The circuit is associated with the unitary $N$-matrix that
transforms the photon creation operators on its $N$ edges. The computer then
calculates the squared moduli of the matrix elements in the $n$-photon
representation of the unitary matrix on the $N$-mode bosonic Fock space, by
repeatedly entering $n$ photons on the input edges of the circuit and
measuring the frequencies of the photons on the output edges.

Circuits are constructed from beam splitters by spatial composition,
physically connecting two circuits at their shared edges, and temporal
composition, repeatedly executing the circuit over consecutive time-bins.
These compositions generate representations for unitary matrices from the
unitary matrices in two dimensions. Physical limitations that challenge the
implementation of this theoretical scheme include:

\begin{itemize}
\item The scheme depends on the ability to create photons on demand on the
input edges and accurately count photons on the output edges.

\item Photons on the internal edges may be lost to or gained from the
external environment.

\item Beam splitters are not precisely configurable, and this introduces
noise into the algorithm.
\end{itemize}

Near-term applications of the photonic computer exploit the algorithmic
complexity that arises from the exponential growth in the dimension of the $n
$-photon representation of $\mathsf{U}[N]$ as the number of photons is
increased. In the boson sampling scheme, the circuit is programmed by the
configurations of the beam splitters at each time-bin, and the distribution
of the output photon counts is estimated by repeated execution. Including a
classical optimisation loop performed outside the quantum calculation, this
hybrid scheme has been proposed as an approach to quantum machine learning.

The near-term ambitions of photonic computing are curtailed by the lack of
interaction between photons, preventing the full implementation of the
universal quantum computer. Long-term applications resolve this shortfall by
introducing additional protocols, including non-deterministic algorithms
that utilise measurement of ancillary photons as operations on the internal
state, and dynamic configuration of beam splitters in response to
measurements of the output state. Implementation of these protocols adds
theoretical and physical challenges to the development of the photonic
computer.

\section{Literature review}

The application of category theory in quantum mechanics has a long history,
with both fields developing in parallel through the twentieth century. The
standard introduction to category theory is the book \cite{MacLane1978}. The
modern approach to symmetric monoidal categories utilises string diagrams,
with comprehensive but readable introductions in the books \cite%
{Coecke2017,Heunen2019}. These texts also present the ZX calculus, a
theoretical cornerstone of the measurement-based approach to quantum
computing. The recent paper \cite{Felice2022} represents the ZX calculus on
the bosonic state space, extending the string diagram approach to photonic
computation.

Quantum computing recognises that quantum mechanics can be utilised to
perform complex calculations. The history of this development is summarised
in the book \cite{Nielsen2010}, which also introduces the high-profile
algorithms that demonstrate quantum advantage and considers their
implementations in real quantum systems. Developments in measurement-based
quantum computing, the leading paradigm for photonic computation, are
reviewed in the articles \cite{Briegel2009,Browne2016}.

Quantum information processing in photonic systems is outlined in the book 
\cite{Kok2010} and the articles \cite{Kok2007,Kok2007a,Kok2016}. Beam
splitters and phase shifters are the basic physical gates used for the
construction of photonic circuits, and there is a large body of literature
that discusses this; see for example \cite%
{Fearn1987,Campos1989,Kim2001,Yarnall2007,Tan2018}. The use of projective
measurement to implement multi-qubit operations is introduced in the
articles \cite{Knill2000,Knill2001}, enabling the development of universal
quantum computing on the photonic computer.

The focus in this essay is motivated by conversations with the team at Orca
Computing, and more details on specific topics can be found in articles by
the team and their collaborators. Time-bin approaches are considered in \cite%
{Humphreys2013,Humphreys2014}. The parity map is analysed in detail in the
article \cite{Bradler2021}, and this is used to develop novel quadratic
optimisation methods as potential near-term applications of photonic
computing. There is a rich and developing body of literature covering
quantum machine learning and similar classical-quantum hybrid algorithms
proposed as noisy intermediate-scale applications; see for example \cite%
{Dunjko2016,Havlicek2018,Benedetti2019,Glasser2019,Romero2019,Huang2021,Nguyen2021,Bharti2022,Gao2022}%
. These approaches exploit quantum mechanics without necessarily requiring
the full power of universal quantum computing, with a greater tolerance for
the errors in near-term architectures.

\end{document}